\documentclass[AMA,STIX1COL]{WileyNJD-v2}
\usepackage{moreverb,url}
\usepackage{booktabs}
\usepackage{multirow}
\usepackage{amsmath, amsthm}
\usepackage{subfigure}
\newcommand\independent{\protect\mathpalette{\protect\independenT}{\perp}}
\def\independenT#1#2{\mathrel{\rlap{$#1#2$}\mkern2mu{#1#2}}}

\newtheorem{prop}{Proposition}

\newcommand\BibTeX{{\rmfamily B\kern-.05em \textsc{i\kern-.025em b}\kern-.08em
T\kern-.1667em\lower.7ex\hbox{E}\kern-.125emX}}

\articletype{Article Type}%

\received{<day> <Month>, <year>}
\revised{<day> <Month>, <year>}
\accepted{<day> <Month>, <year>}

\begin{document}

\title{Group sequential two-stage preference designs}

\author{Ruyi Liu$^1$, Fan Li$^1$, Denise Esserman$^1$, Mary M. Ryan$^{2,*}$}




\authormark{LIU \textsc{et al}}

\address[1]{\orgdiv{Department of Biostatistics}, \orgname{Yale School of Public Health}, \orgaddress{\state{New Haven, Connecticut}, \country{USA}}}

\address[2]{\orgdiv{Departments of Population Health Sciences \& Biostatistics and Medical Informatics}, \orgname{University of Wisconsin-Madison}, \orgaddress{\state{Madison, Wisconsin}, \country{USA}}}

\corres{*Mary M. Ryan, Departments of Population Health Sciences \& Biostatistics and Medical Informatics, University of Wisconsin-Madison, Madison, WI 53726, USA. \email{mary.ryan@wisc.edu}}

\abstract[Abstract]{The two-stage preference design (TSPD) enables the inference for treatment efficacy while allowing for incorporation of patient preference to treatment. It can provide unbiased estimates for selection and preference effects, where a selection effect occurs when patients who prefer one treatment respond differently than those who prefer another, and a preference effect is the difference in response caused by an interaction between the patient’s preference and the actual treatment they receive. One potential barrier to adopting TSPD in practice, however, is the relatively large sample size required to estimate selection and preference effects with sufficient power. To address this concern, we propose a group sequential two-stage preference design (GS-TSPD), which combines TSPD with sequential monitoring for early stopping. In the GS-TSPD, pre-planned sequential monitoring allows investigators to conduct repeated hypothesis tests on accumulated data prior to full enrollment to assess study eligibility for early trial termination without inflating type I error rates. Thus, the procedure allows investigators to terminate the study when there is sufficient evidence of treatment, selection, or preference effects during an interim analysis, thereby reducing the design resource in expectation. To formalize such a procedure, we verify the independent increments assumption for testing the selection and preference effects and apply group sequential stopping boundaries from the approximate sequential density functions. Simulations are then conducted to investigate the operating characteristics of our proposed GS-TSPD compared to the traditional TSPD. We demonstrate the applicability of the design using a study of Hepatitis C treatment modality.}

\keywords{Two-stage randomized clinical trial, group sequential monitoring, independent increments, preference effect, preference trial}

\maketitle

\section{Introduction} \label{section:Intro}
\begin{figure}[htbp]
    \centering
    \subfigure{\includegraphics[width=0.8\textwidth]{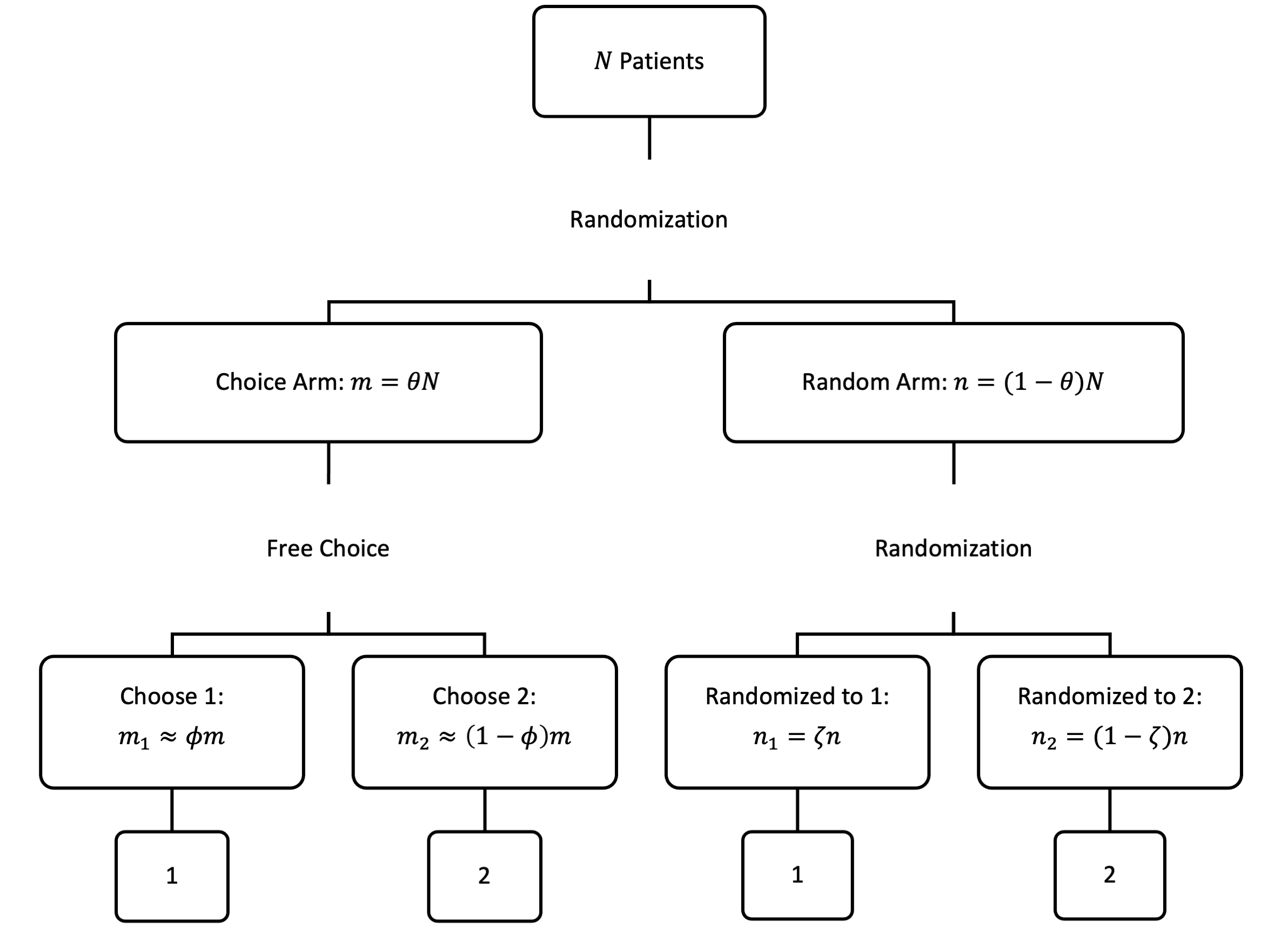}} 
    \caption{Schematic illustration of the traditional two-stage preference design, where $\theta$ is the first-stage randomization ratio to the choice arm, $\phi$ is the preference rate for treatment $1$, and $\zeta$ is the second-stage randomization ratio to treatment $1$.}
    \label{fig: traditional two-stage randomized preference trial}
\end{figure}

Traditional randomized controlled trials (RCTs) are typically designed to examine the efficacy or effectiveness of a treatment in the study population; they cannot, however, account for the impact an individual's treatment preference may have on the individual's response.\cite{cameron_extensions_2018} To incorporate patient preference for treatment, R\"ucker \cite{rucker_two-stage_1989} proposed a two-stage preference design (TSPD), a combination of a randomized and non-randomized design (see Figure \ref{fig: traditional two-stage randomized preference trial} for a schematic illustration), where in the first stage all participants are randomized to a choice arm or a random arm and in the second stage, participants in the random arm are randomly assigned to treatment conditions while patients in the choice arm are allowed to choose their preferred treatment.\cite{turner_sample_2014} This design has also been generalized to accommodate stratification\cite{cameron_sample_2018} and non-random assignment in the first stage.\cite{wang_design_2022} Typically, the TSPD allows the separation of the selection and preference effects from the treatment effect. A selection effect refers to the expected difference in response between individuals who prefer and would choose one treatment (if allowed to do so) over the other treatment;
a preference effect is the difference in response induced by an interaction between the patient's treatment preference and the actual treatment received. However, estimating these effects with sufficient power almost always requires a larger sample size than a traditional RCT targeting the treatment effect. This may be a potential limitation for adopting TSPD in practice, especially when the study is constrained by financial and logistical resources. 

In this article, we propose the use of sequential monitoring techniques for a TSPD to reduce the expected sample size. A group sequential design performs repeated significance tests on the data as it accumulates.\cite{pocock_group_1977} If adequate evidence is found to reject the null hypothesis prior to complete enrollment, the study may be terminated early. Compared to a fixed sample design, a group sequential design may reduce study resources when sufficient evidence of the primary objective is identified during an interim analysis. It is also ethically attractive because early termination can prevent unnecessary experimentation on participants for treatments with established benefits. 
\cite{kim_independent_2020} Sequential monitoring procedures have been adapted in simple randomized controlled trials for a variety of responses, including continuous, binary, and time-to-event outcomes.\cite{jennison_group-sequential_1997} Here, we will consider the application of sequential monitoring methods to TSPD with continuous and binary outcome measures, and formally assess whether such methods can lead to a meaningful reduction in resource requirements for designs accommodating patient treatment preference. 

Our study is motivated by the design of a trial that studies the best care delivery for treatment of Hepatitis C (HCV). 
Infection with HCV is the leading cause of cirrhosis, hepatocellular carcinoma, and liver transplantation\cite{razavi_chronic_2013}, as well as a major cause of morbidity and mortality in the United States.\cite{rosenberg_prevalence_2018} In addition, HCV incidence has increased in recent years, more than doubling since 2013, with particular risk among injection drug users.\cite{CDC_2020_2022} Many infected individuals, particularly drug and alcohol users, do not seek treatment or have been denied treatment for a variety of factors including disease and drug-use related stigma and systematic barriers.\cite{cameron_sample_2018, trooskin_we_2020} According to national guidelines, HCV should be treated in a specialty hospital clinic. For patients who are unable or reluctant to obtain care in a traditional hospital setting, however, mobile medical clinics (MMCs) have emerged to support more patients seeking treatment. Notifying patients of their healthcare settings, investigators expected that their preference for where to obtain their treatment would meaningfully impact health-related quality of life (HRQoL) and the ultimate cure rate. Besides, patients receiving care at MMCs might have better outcomes overall. Thus, R\"ucker's two-stage design could be useful to evaluate the treatment effect and the impact of patient preference on these outcomes.\cite{rucker_two-stage_1989} To detect preference and selection effects in this trial, the required sample size under a TSPD was not trivial. A stratified two-stage procedure was previously developed \cite{cameron_sample_2018, cameron_extensions_2018} to account for differential preference rates, which can reduce the required sample size to some extent. The utility of group sequential methods in this setting, however, remains unexplored. 

The remainder of the article is structured as follows. Section \ref{section:background} provides a review of traditional TSPD and group sequential monitoring methods, including assumptions and notation. In Sections \ref{subsection:design}-\ref{subsection:samplesize}, we propose a new group sequential two-stage preference design (GS-TSPD) for Normally distributed continuous responses, and justify the applicability of sequential monitoring to selection and preference effects by establishing the independent increments assumption for these effects. Section \ref{subsection:binar} extends our GS-TSPD to binary responses and presents the relevant design considerations. In Section \ref{section:sims}, we conduct simulation studies to validate our proposed GS-TSPD with continuous and binary outcome responses. Illustrative data examples are presented in Section \ref{section:example}. Finally, Section \ref{section:discussion} concludes with a discussion.  

\section{Background} \label{section:background}
\subsection{Two-stage preference design}
Researchers can utilize TSPDs to better understand the impact of treatment preference on treatment response through estimation of selection and preference effects. In a TSPD, participants follow a two-stage randomization procedure (Figure \ref{fig: traditional two-stage randomized preference trial}). Assume we have $N$ participants, and let $\theta$ represent the proportion of participants randomized to the choice arm, such that we have $m=\theta N$ participants assigned to the choice arm. We assume all participants have a treatment preference and there are no undecided or ambivalent participants. Let $\phi$ represent the expected preference rate for treatment $1$, i.e., the expected proportion of participants who prefer treatment $1$ over treatment $2$, such that $m_1 = \phi m$ and $m_2 = (1-\phi)m$ are the sample sizes for those choosing treatment $1$ and $2$, respectively. Similarly, let the number of participants randomized to the random arm be $n=(1-\theta) N$, with $n_1 = \zeta n$ and $n_2 = (1-\zeta) n$ participants randomly assigned to treatment $1$ and $2$, respectively, where  $\zeta$ is the proportion randomly assigned to treatment $1$. 

First let us consider the case where $y_{ijk}$ is a continuous response for participant $k$ assigned treatment $i$ $(i = 1,2)$ who prefers treatment $j$ $(j=1,2)$. Following R\"ucker,\cite{rucker_two-stage_1989} the outcome can be modeled as
\begin{equation}
\label{eq: normal continuous model} 
y_{i j k}=\mu+\tau_{i}+\nu_{j}+\pi_{i j}+\epsilon_{i j k},
\end{equation}
where $\mu$ is the overall mean response, $\tau_{i}$ is the treatment effect of treatment $i$, $\nu_{j}$ is the selection effect of preference $j$, $\pi_{i j}$ is the preference effect for treatment $i$ and preference $j$, and $\epsilon_{i j k}$ is the random error term that follows a normal distribution, i.e., $\epsilon_{i j k} \sim \mathcal{N}(0,\sigma^2_c)$. In addition, we note that model \eqref{eq: normal continuous model} is essentially a two-way analysis of variance model under effect coding, and that the following restrictions are required for identifiability: (i) $\tau_1+\tau_2=0$; (ii) $\phi\nu_1+(1-\phi)\nu_2=0$; (iii) $\phi\pi_{i1}+(1-\phi)\pi_{i2}=0$ for $i=1,2$; (iv) $\pi_{1j}+\pi_{2j}=0$ for $j=1,2$.

Table \ref{tab:definitions} provides an explicit characterization of the group structure, proportion and mean response within each group defined by the arm and treatment preference. We define $\mu_{1}$ and $\mu_{2}$ as the mean response for participants receiving treatment 1 and treatment 2, respectively, in the random arm. These parameters can be directly estimable from the observed data by sample means. Similarly, we define $\mu_{ij}$ as the mean response for participants receiving treatment $i$ ($i=1, 2$) and preferring treatment $j$ ($j=1,2)$. Of note, it is possible to directly estimate $\mu_{11}$ and $\mu_{22}$ from the observed data under the choice arm. However, the parameters $\mu_{12}$ and $\mu_{21}$ need to be indirectly inferred. Due to the first-stage randomization, the random arm consists of a mixture of participants who received their preferred treatment and who did not receive their preferred treatment. This assumption allows us to express
\begin{align*}
   \mu_{1}&=\phi\mu_{11}+(1-\phi)\mu_{12},\\
   \mu_{2}&=\phi\mu_{21}+(1-\phi)\mu_{22},
\end{align*}
which then gives
\begin{align*}
   \mu_{12}&=(\mu_1-\phi\mu_{11})/(1-\phi),\\
   \mu_{21}&=(\mu_2-(1-\phi)\mu_{22})/\phi.
\end{align*}
By definition of analysis of variance model \eqref{eq: normal continuous model} along with the set of constraints, we can write
\begin{align*}
\mu=&\frac{\mu_{1}+\mu_{2}}{2},~~~~~~~~~~~~~~~~~~~~~~~~~~~~~~~~~~~~~
\tau_{i}=\mu_{i}-\mu,~~~i\in\{1,2\},\\
\nu_{j}=&\frac{\mu_{1j}+\mu_{2j}}{2}-\mu,~~~j\in\{1,2\},~~~~
\pi_{ij}=\mu_{ij}-\tau_{i}-\nu_{j}-\mu,~~~i,j\in\{1,2\}.
\end{align*}
Based on these explicit relationships, we can express the differences in treatment, selection, and preference effects as follows:
\begin{align*}
  \Delta\tau=&\tau_{1}-\tau_{2}=\mu_{1}-\mu_{2},\\
  \Delta\nu=&\nu_{1}-\nu_{2}=\frac{(\mu_{11}-\mu_{12})+(\mu_{21}-\mu_{22})}{2},\\
  \Delta\pi=&\frac{(\pi_{11}+\pi_{22})-(\pi_{12}+\pi_{21})}{2}=\frac{(\mu_{11}+\mu_{22})-(\mu_{12}+\mu_{21})}{2},
\end{align*}
which are the target parameters of interest. 
Finally, the selection and preference effects can be further expressed as functions of the observable mean responses as
\begin{align*}
\Delta\nu&=\frac{\phi(\mu_{11}-\mu_1)-(1-\phi)(\mu_{22}-\mu_2)}{2\phi(1-\phi)}\\
\Delta\pi&=\frac{\phi(\mu_{11}-\mu_1)+(1-\phi)(\mu_{22}-\mu_2)}{2\phi(1-\phi)}.
\end{align*}

\begin{table}[htbp]
\caption{Group structure and mean response in each group under a two-stage preference design. In the table, $\phi$ is the preference rate for treatment 1, and $\zeta$ is the probability for randomizing to receive treatment 1. The mean parameter in the square brackets are unobserved quantities under the random arm.}
\label{tab:definitions}
\centering
\begin{tabular}{ccccc}
\toprule
 & \multicolumn{2}{c}{Choice arm} & \multicolumn{2}{c}{Random arm}\\
\midrule 
 Received treatment & Prefer 1 & Prefer 2 & Prefer 1 & Prefer 2\\
 \midrule 
 Receive 1 & $\phi$ & $-$ & $\phi\zeta$ & $(1-\phi)\zeta$\\
 & & & $[\mu_{11}]$ & $[\mu_{12}]$\\
Observable mean parameter & $\mu_{11}$ & $-$ & \multicolumn{2}{c}{$\mu_1$}\\
 \midrule
 Receive 2 & $-$ & $1-\phi$ & $\phi(1-\zeta)$ & $(1-\phi)(1-\zeta)$ \\
 & & & $[\mu_{21}]$ & $[\mu_{22}]$\\
Observable mean parameter & $-$ & $\mu_{22}$ & \multicolumn{2}{c}{$\mu_2$}\\
 \bottomrule
\end{tabular}
\end{table}

Motivated by the above relationships between model parameters and mean responses, R\"ucker previously derived the test statistics for the treatment, selection, and preference effects in a fixed sample TSPD.\cite{rucker_two-stage_1989} The treatment effect test statistic, ${\mathcal{T}}_{\tau}$, is the standardized difference in means for the random arm; the selection effect test statistic, ${\mathcal{T}}_{\nu}$, is the standardized difference of the additional increase in response caused by selecting treatment $1$ as opposed to treatment $2$; and the test statistic for the preference effect, ${\mathcal{T}}_{\pi}$, is the standardized sum of the additional increase in response caused by receiving the preferred treatment:
$${\mathcal{T}}_{\tau}=\frac{\bar{y}_{1}-\bar{y}_{2}}{\sqrt{{\widehat{\operatorname{Var}}\left(y_{1}\right)}/{n_{1}}+{\widehat{\operatorname{Var}}\left(y_{2}\right)}/{n_{2}}}}$$
$${\mathcal{T}}_{\nu}=\frac{z_{1}-z_{2}}{\sqrt{\widehat{\operatorname{Var}}\left(z_{1}-z_{2}\right)}}$$
$${\mathcal{T}}_{\pi}=\frac{z_{1}+z_{2}}{\sqrt{\widehat{\operatorname{Var}}\left(z_{1}+z_{2}\right)}},$$
where $\bar{y}_{1}$ is the average observed response of all participants in random arm randomized to treatment $1$, i.e., $\bar{y}_{1} = \frac{1}{n_1}\sum_{h=1}^{n_{1}} y_{1h}$, $\bar{y}_{2}$ is the average response of all participants in random arm randomized to treatment $2$, i.e., $\bar{y}_{2} = \frac{1}{n_2}\sum_{h=1}^{n_{2}} y_{2h}$, $z_{1}=\left(\sum_{h=1}^{m_{1}} x_{1 h}\right)-m_{1} \bar{y}_{1}$ and   
$z_{2}=\left(\sum_{h=1}^{m_{2}} x_{2 h}\right)-m_{2} \bar{y}_{2}$, and $x_{1h}, x_{2h}, y_{1h}, y_{2h}$ are used to denote the response of a participant in the choice arm choosing treatment $1$, that of a participant in the choice arm choosing treatment $2$, that of a participant in the random arm randomized to treatment $1$, and that of a participant in the random arm randomized to treatment $2$, respectively. Note that the treatment effect test statistic only uses information from participants in the random arm, while the selection and preference effect test statistics use information from participants in both the choice and random arms.

For a binary outcome, we consider the linear probability model developed in Cameron et al.\cite{cameron_extensions_2018} where 
$p_{ijk}$ is the mean outcome of participant $k$ treated with treatment $i$ $(i = 1,2)$ who prefers treatment $j$ $(j=1,2)$ and is represented by
\begin{equation}
\label{eq: binomial model} 
p_{i j k}=p+\tau_{i}+\nu_{j}+\pi_{i j}.
\end{equation}
Here $p$ is the overall response proportion, and $\tau_{i}$, $\nu_{j}$, and $\pi_{i j}$ are defined analogously as in model \eqref{eq: normal continuous model}. The observed binary outcome $y_{ijk}\sim\text{Bernoulli}(p_{ijk})$. The detailed testing procedure is in general similar to that for the continuous response, but recognizes the binomial nature of the response (hence the variance of the response is a function of its mean); we refer Cameron et al.\cite{cameron_extensions_2018} for full details with binary responses.

\subsection{Group sequential design}
A group sequential monitoring approach can be used to determine whether a study can be terminated prior to reaching full enrollment. Under this approach, repeated significance tests (interim analyses) are conducted based on all available data at the time of analysis, often at pre-specified time points, with a maximum of $L$ interim analyses. Between consecutive analyses, a specified number of additional participants are enrolled and evaluated. If the test statistic's magnitude at an analysis $l$ does not meet the interim analysis's significance threshold (stopping boundary), the trial continues; otherwise the study is terminated early and no additional data are collected. Critical factors to be considered while designing a group sequential trial are the timing, number of the interim analyses and the choice of stopping boundary, all of which can affect the design's operating characteristics.\cite{grayling_accounting_2021} If we are unsuccessful in terminating a trial at a given interim analysis, the next analysis will include all prior data plus newly accumulated data, resulting in successive test statistics which may not be independent. This repeated testing results in an inflation of the overall type I error rate if these stopping boundaries are left unadjusted from the usual $z_{1-\alpha/2}$ level. Therefore, to maintain our desired overall type I error rate, we need to employ one of the group sequential design (GSD) families, a set of parameterized boundary functions that link the successive analyses to the data accumulated and the rejection of the null hypothesis. Different boundary functions allow the probability of stopping to vary with time differently.\cite{emerson_frequentist_2007} For simplicity, we only consider stopping early for efficacy in this article. Similar methods can be used when stopping for futility is also required. 

Additionally, keeping all other factors the same, when we have more possible interim analyses, i.e., larger $L$, the number of participants, events, or incremental information gained between analyses is smaller and there are more opportunities to assess the accrued data for early termination. On the other hand, a larger number of interim analyses would require more drastic adjustments to interim stopping boundaries compared to fixed sample decision thresholds, further requiring a larger maximum sample size at analysis $L$. This means that in the worst-case scenario of continuing to the end of a group sequential study, more interim analyses would require more samples. Although a larger $L$ is associated with a larger maximum sample size, it can concurrently reduce the probability of reaching an analysis with a sample size greater than the fixed sample design, thus increasing the likelihood of saving samples for a particular study.

By adjusting the stopping boundaries we compare our test statistic to during interim tests, group sequential methods ensure that the overall type I error rate is controlled at the desired level.\cite{demets_interim_1994} 
Here, we consider pre-specified alpha-spending functions to calculate the stopping boundaries. These functions restrict the overall type I error rate while enabling flexibility in the number and timing of interim analyses by specifying a sequence of nominal significance levels to be divided up at each interim analysis based on the amount of information available, with the total adding up to the predetermined overall significance level $\alpha$.\cite{demets_interim_1994} Alpha-spending functions are often formulated in relation to the information fraction at the $l^{th}$ analysis, $\Pi_l$, defined as the ratio of the variance at analysis $l$ over the variance at final analysis $L$.\cite{lan1995sequential} For many models, including those considered here, this is proportional to the ratio of the current sample size over the maximal planned sample size if we assume constant variance over the entire study population.

\begin{figure}[h]
    \centering
    \subfigure{\includegraphics[width=0.6\textwidth]{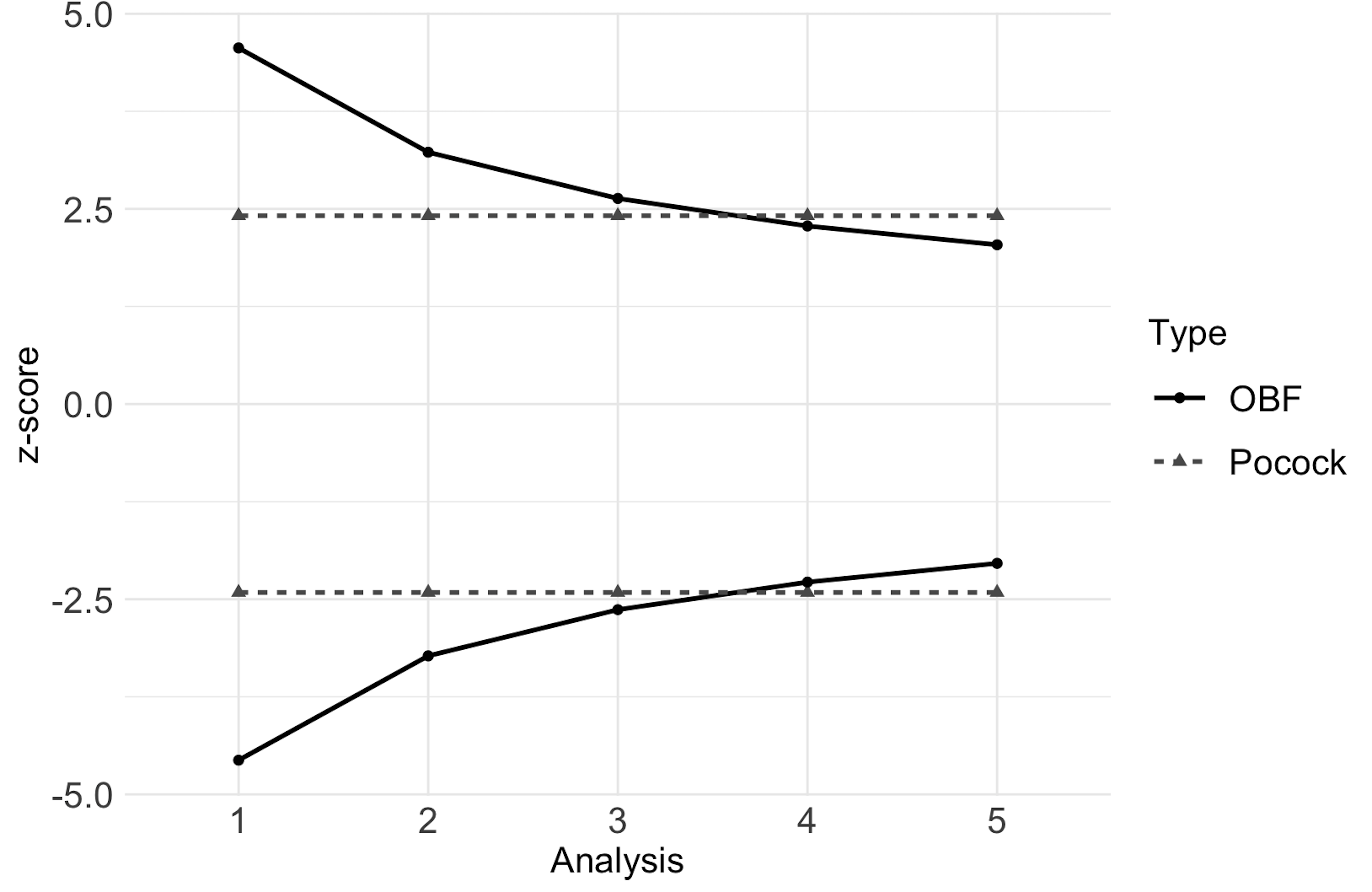}} 
    \caption{Pocock and O'Brien-Fleming (OBF) stopping boundaries on the $z$-statistic scale with a continuous Normally distributed endpoint, a maximum of $L=5$ analyses, and assuming a two-sided type I error of $5$\%.}
    \label{fig: stopping boundaries}
\end{figure}

We will specifically focus on two alpha-spending functions, the Pocock alpha-spending function\cite{lan_group_1989} 
 \begin{equation}
\label{eq: Pocock alpha-spending}  \alpha^*\left(\Pi_l\right)=\alpha \log \left(1+(e-1) \Pi_l\right)
\end{equation}

 \noindent
 and the O'Brien-Fleming (OBF) alpha-spending function\cite{obrien_multiple_1979} 

 \begin{equation}
\label{eq: OBF alpha-spending} 
\alpha^*\left(\Pi_l\right)=2\left(1-\Phi\left(\frac{z_{{\alpha}/{2}}}{\sqrt{\Pi_l}}\right)\right),
\end{equation}

\noindent
where $\alpha^*\left(\Pi_l\right)$ denotes the significance level applied at the $l^{th}$ analysis
. Each alpha-spending function ``spends'' the nominal type I error differently across analyses. The Pocock alpha-spending function, for example, has a large initial then decreasing $\alpha$ expenditure while the OBF alpha-spending function spends a small amount initially with increasing expenditure with each interim analysis. Alpha expenditures at each analysis can subsequently be translated into stopping boundaries on more interpretable scales, such the $z$-statistic scale or the scale of the treatment effect, against which we may compare our observed data. For illustration, Figure \ref{fig: stopping boundaries} compares the stopping boundaries on the $z$-statistic scale given by the Pocock and OBF alpha-spending functions in a two-sided test assuming five total analyses equally spaced over time. Pocock boundaries are characterized by a constant p-value threshold at each analysis, leading to constant boundaries on $z$-statistic scale; as data is accrued, this translates to a smaller portion of type I error spent at each subsequent interim analysis. Meanwhile, the OBF alpha-spending function increases expenditure in the nominal significance levels as the study moves through interim analyses, leading to progressively tightening boundaries on $z$-statistic scale. This means that the probability of stopping at the first analysis under an OBF design is smaller than under a Pocock design. The trade-off for this early conservativeness in early stopping is a larger expected sample size than that under the Pocock design, but a smaller maximum sample size at analysis $L$.

While not explored in this work, it is important to note that other alpha-spending functions exist, including formulations that are entirely user-defined, that will produce stopping boundaries of different shapes than those illustrated in Figure \ref{fig: stopping boundaries} while still maintaining type I error. For example, Haybittle-Peto\cite{haybittle_repeated_1971, peto_design_1976} boundaries are far more conservative than either Pocock or OBF boundaries, making the probability of early stopping very small unless the observed effect size is very large; this approach may be useful when other research priorities such as gathering safety data are a major concern, but provides few advantages in terms of expected savings in sample size. In the existing literature, sequential monitoring has been most commonly used in randomized clinical trials of therapeutic agents, which may have different research priorities and considerations than more pragmatic or patient-centered studies. Pragmatic trials are often designed to be more flexible and adaptive, whereas therapeutic trials are generally designed to have larger sample sizes in order to detect rare adverse events should they occur. Therapeutic trials also have a strong inclination to promptly terminate the study if any initial indications of harm emerge, which is of less concern in many pragmatic studies investigating lower risk interventions. Thus, investigators of pragmatic trials should be aware that published advice on alpha-spending function choice may not align with the overarching goals of their specific study, and they should carefully consider their unique objectives when choosing an appropriate alpha-spending function. 

Alpha-spending functions will generate valid boundary thresholds that maintain type I error if the sequential sampling density of the test statistics is approximately multivariate Normal with an independent increments covariance structure, known as the \emph{independent increments property}. Independent increments assume that the change in cumulative test statistics resulting from the accumulation of new data is independent of the cumulative test statistics in previous interim analyses. By verifying this assumption for the cumulative test statistics, an implicit recursive formula is defined to compute the sequential sampling density.\cite{armitage_repeated_1969} The independent increments assumption has been established for treatment effects in traditional RCTs.\cite{kim_independent_2020} This assumption has not be verified for the selection and preference effect in the two-stage preference design framework; the work in the following section addresses this issue.

\section{Group sequential two-stage preference design with continuous responses} \label{section:model}
\subsection{Study design} \label{subsection:design}
\begin{figure}[h]
    \centering
    \subfigure{\includegraphics[width=1\textwidth]{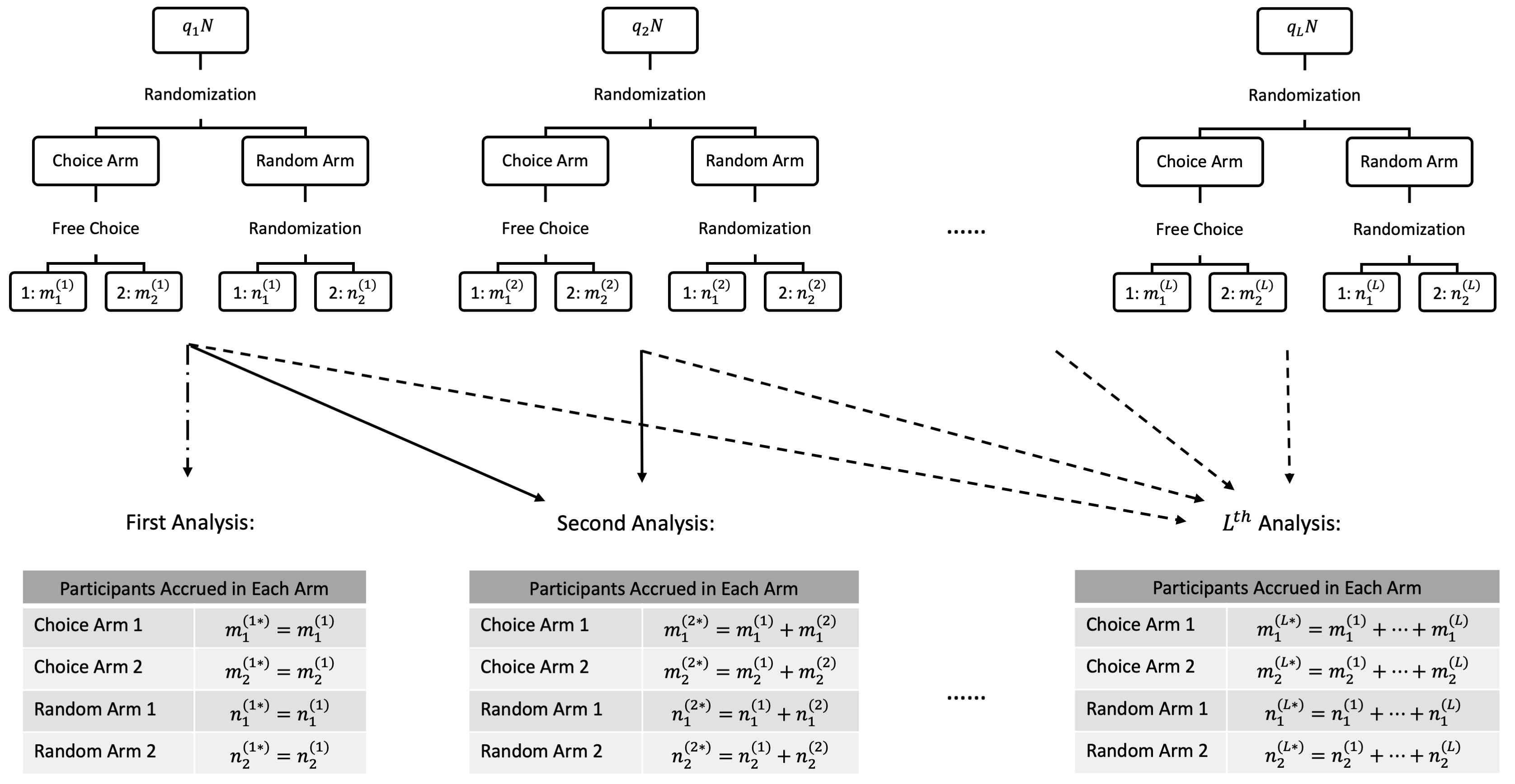}} 
    \caption{Schematic of a Group Sequential - Two-Stage Preference Design (GS-TSPD) assuming all participants have a treatment preference. For all data available at $l^{th}$ analysis ($l = 1,\cdots,L$), where $L$ is the maximum number of analyses, $m_{1}^{(l*)}, m_{2}^{(l*)}, n_{1}^{(l*)}, n_{2}^{(l*)}$ denote the number of participants in the choice arm choosing treatment $1$, the number of participants in the choice arm choosing treatment $2$, the number of participants in the random arm assigned treatment $1$, and the number of participants in the random arm assigned treatment $2$, respectively. Without the asterisk in the superscript, the quantity only accounts for the newly accrued data.}
    \label{fig: group sequential two-stage preference design}
\end{figure}

To leverage the cost effective property of a group sequential design in studying selection and preference effects from the TSPD, we propose a group sequential two-stage preference design (GS-TSPD). We assume there are at most $N$ participants with at most $L$ analyses conducted. Let $q_1,\ldots,q_L$ denote the proportion of information newly accrued at each analysis, such that $\sum_{l=1}^L q_l = 1$. If after interim analysis $l-1$ ($l = 2,\cdots,L$), we do not have sufficient evidence to stop the trial for efficacy, we will recruit an additional $q_{l}N$ participants prior to conducting the $l^{th}$ analysis. A schematic illustration of the GS-TSPD is shown in Figure \ref{fig: group sequential two-stage preference design}. For simplicity and without loss of generality, we assume the second-stage treatment randomization ratio $\zeta$ is $0.5$. Similar to the fixed sample TSPD setting, $\theta$ represents the proportion of participants randomized to the choice arm and and $\phi$ represents the preference rate for treatment $1$. Let $m^{(l)}$ and $n^{(l)}$ represent the total number of new participants randomized to the choice and random arms at analysis $l$, respectively. Thus, $m_{1}^{(l)}, m_{2}^{(l)}, n_{1}^{(l)}, n_{2}^{(l)}$ represent the number of participants randomized to the choice arm who choose treatment $1$, the number of participants randomized to the choice arm who choose treatment $2$, the number of participants randomized to the random arm assigned treatment $1$, and the number of participants randomized to the random arm assigned treatment $2$, respectively, such that  $m_1^{(l)} + m_2^{(l)} = m^{(l)}$, and $n_1^{(l)} + n_2^{(l)} = n^{(l)}$. 

\subsection{Test statistics of analysis} \label{subsection:stats}

Following model \eqref{eq: normal continuous model} and assuming a continuous outcome, we use $x_{1h}^{(l)}, x_{2h}^{(l)}, y_{1h}^{(l)}, y_{2h}^{(l)}$ to denote the responses for newly accrued participant $h$ at the $l^{th}$ analysis in the choice arm choosing treatment $1$, in the choice arm choosing treatment $2$, in the random arm assigned treatment $1$, and in the random arm assigned treatment $2$, respectively. The superscript `$l$' notation is used to refer to information accrued between the $(l-1)^{th}$ and $l^{th}$ interim analyses, while the superscript `$l$*' notation is used to refer to the all information available at $l^{th}$ interim analysis. For example, $n_1^{(l*)} = n_1^{(1)} + \cdots + n_1^{(l)}$. 

In a fixed sample TSPD, data are analyzed once at the end of the study, when all data have been collected; however, in the GS-TSPD, we test the data sequentially, with each test statistic that we compare against stopping boundaries calculated using all information available at the time of the interim analysis -- we will refer to this as the ``test statistic of analysis''. Note that our proposed GS-TSPD considers the simple scenario where we assume one of the three effects (treatment, selection, preference) is of primary interest. Thus, we develop the methodology in parallel for each of the effects. The test statistics of analysis for three estimands at the $l^{th}$ interim analysis are calculated as follows, using the same formulae as in fixed sample TSPD but now based on the cumulative information at the $l^{th}$ analysis:

\begin{equation}
\label{eq: treatment effect test stats} 
{\mathcal{T}}_{\tau}(l\text{*}) = \frac{\overline{y_{1}^{(l*)}}-\overline{y_{2}^{(l*)}}}{\sqrt{{\widehat{\operatorname{Var}}\left(y_{1}^{(l*)}\right)}/{n_{1}^{(l*)}}+{\widehat{\operatorname{Var}}\left(y_{2}^{(l*)}\right)}/{n_{2}^{(l*)}}}}
\end{equation}

\begin{equation}
\label{eq: selection effect test stats} 
{\mathcal{T}}_{\nu}(l\text{*}) = \frac{z_{1}^{(l*)}-z_{2}^{(l*)}}{\sqrt{ \widehat{\operatorname{Var}}\left(z_{1}^{(l*)}-z_{2}^{(l*)}\right)}}
\end{equation}

\begin{equation}
\label{eq: preference effect test stats} 
{\mathcal{T}}_{\pi}(l\text{*}) = \frac{z_{1}^{(l*)}+z_{2}^{(l*)}}{\sqrt{ \widehat{\operatorname{Var}}\left(z_{1}^{(l*)}+z_{2}^{(l*)}\right)}},
\end{equation}

where
$$z_{1}^{(l*)}=\left(\sum_{h=1}^{m_{1}^{(l*)}} x_{1 h}^{(l*)}\right)-m_{1}^{(l*)} \overline{{y}_{1}^{(l*)}}$$
$$z_{2}^{(l*)}=\left(\sum_{h=1}^{m_{2}^{(l*)}} x_{2 h}^{(l*)}\right)-m_{2}^{(l*)} \overline{y_{2}^{(l*)}} $$
We explicitly define the estimated variances in the denominators of these test statistics of analysis in Appendix 1. With Equations \eqref{eq: treatment effect test stats}, \eqref{eq: selection effect test stats}, \eqref{eq: preference effect test stats}, the group sequential test statistics of analysis can be computed for the primary effect of interest. The trial will be terminated at analysis $l$ if the interim test statistic of analysis lies outside of the stopping boundaries provided by the chosen alpha-spending function; the trial will continue if the interim test statistic of analysis lies within the stopping boundaries and the process will be repeated until we exceed the interim stopping boundaries or reach the final analysis with the pre-specified total sample size.

\subsection{Sequential sampling density and independent increments} \label{subsection:seq}
In order to establish the properties of a GS-TSPD, we introduce what we will refer to as the ``cumulative test statistics'', $S_l$, for each of the three effects at the $l^{th}$ analysis:
$$\text{Treatment Difference:}\quad S_{l, \tau} = \sum_{r=1}^{l} {\mathcal{T}}_{\tau}(r) =\sum_{r=1}^{l} \frac{\overline{y_{1}^{(r)}}-\overline{y_{2}^{(r)}}}{\sqrt{{\widehat{\operatorname{Var}}\left(y_{1}^{(r)}\right)}/{n_{1}^{(r)}}+{\widehat{\operatorname{Var}}\left(y_{2}^{(r)}\right)}/{n_{2}^{(r)}}}} $$
$$\text{Selection Difference:}\quad S_{l, \nu} = \sum_{r=1}^{l} {\mathcal{T}}_{\nu}(r) = \sum_{r=1}^{l}\frac{z_{1}^{(r)}-z_{2}^{(r)}}{\sqrt{\widehat{\operatorname{Var}}\left(z_{1}^{(r)}-z_{2}^{(r)}\right)}} $$
$$\text{Preference Difference:}\quad S_{l, \pi} = \sum_{r=1}^{l} {\mathcal{T}}_{\pi}(r) = \sum_{r=1}^{l}\frac{z_{1}^{(r)}+z_{2}^{(r)}}{\sqrt{\widehat{\operatorname{Var}}\left(z_{1}^{(r)}+z_{2}^{(r)}\right)}}.$$

\noindent
We define the cumulative test statistic as the sum of standardized test statistics over different periods (i.e., within each time period the mean and variance are calculated based on data collected between the two interim analyses), while the test statistics of analysis in Section \ref{subsection:stats} are the standardized test statistics calculated using all prior data.\cite{kim_independent_2020} In general, sampling is terminated the first time $|S_l|$ is larger than upper bound $b_l$, where $b_l$ is the stopping upper bound for the cumulative test statistic at the $l^{th}$ interim analysis. Differences in formulation generally make it easier to establish group sequential properties (such as stopping boundaries) on the cumulative test statistic scale, but once established they may be mapped back to the scale of the test statistic of analysis. Thus, we can equivalently define corresponding $U_l$ and $M_l$ for the upper and lower bounds at the $l^{th}$ analysis, respectively, on the scale of test statistic of analysis, which will be used to determine the termination at $l^{th}$ analysis. 

To obtain boundary statistics from the alpha-spending approach, we need the sequential sampling density, i.e., the joint distribution of the cumulative test statistics. In general, group sequential methods require multivariate integration to calculate the group sequential boundary statistics:
$$\mathbb{P}\left(\left|S_{1}\right| \leq b_{1}, \ldots,\left|S_{L}\right| \leq b_{L}\right)=\int_{-b_{1}}^{b_{1}} \cdots \int_{-b_{L}}^{b_{L}} f\left(s_{1}, \ldots, s_{L}\right) d s_{1} \cdots d s_{L}=1-\alpha$$
where $f$ is the joint density of the cumulative test statistics, and $\alpha$ is the overall significance level.\cite{kim_independent_2020}
However, if we are able to make use of the independent increments assumption for the cumulative test statistics, we may apply an Armitage Normal approximation and reduce the multivariate integration to an implicit recursive formula for the sampling density\cite{emerson_frequentist_2007, kim_independent_2020, armitage_repeated_1969}:  

$$1-\alpha=\ \int_{-b_{L}}^{b_{L}} f_L(s_{L}) d s_{L}$$
$$f_{l}(s)=\int_{-b_{l-1}}^{b_{l-1}} f_{l-1}(u) f_1(s-u) d u,$$
where $f_1$ is the Normal density function and $f_{L}(s)=\int_{-b_{L-1}}^{b_{L-1}} f_{L-1}(u) f_1(s-u) d u$. This allows us to simplify the boundary calculation. If the absolute value of the cumulative test statistic, $|S_l|$, is less than or equal to $b_l$, we will continue the trial and enroll more participants into the study before analysis $l+1$; otherwise, we stop the trial and reject the null hypothesis.

In a GS-TSPD with Normally distributed responses, the increments of the treatment effect cumulative test statistics have been previously proven to be independent.\cite{kim_independent_2020,jennison_group-sequential_1997} This result is asymptotic for non-Normal data as sample size increases, but it remains a good approximation for small samples.\cite{jennison_group_1999} In addition, we have specifically verified that the independent increments assumption also holds for the selection and preference effects -- which use information from both choice and random arms -- in a two-stage preference design framework, as stated in Proposition 1. 
\begin{prop}
\textit{The cumulative test statistic for the selection effect, $S_{l, \nu}$, and preference effect, $S_{l, \pi}$, satisfy the independent increments assumption such that}
$$\operatorname{Cov}\left(S_{l-1,\nu}, S_{l,\nu}-S_{l-1,\nu}\right)=0,~~~\text{and}~~~\operatorname{Cov}\left(S_{l-1,\pi}, S_{l,\pi}-S_{l-1,\pi}\right)=0.$$
\end{prop}
\noindent
Thus, we can guarantee that stopping boundary thresholds generated by traditional alpha-spending functions for the selection and preference effects will maintain the overall type I error rate of the GS-TSPD study. The proof for Proposition 1 is provided in Appendix 2.

\subsection{Design considerations: Inflation factors and maximum GS-TSPD sample size} \label{subsection:samplesize}
When designing a GS-TSPD, we need to establish the primary effect of interest (treatment, selection or preference) and then based on the hypothesized effect size, we can estimate the maximum GS-TSPD sample size -- the sample size if no early termination occurs -- with power $1-\beta$ and an overall significance level $\alpha$. This calculation is based on the required sample size for a corresponding fixed sample TSPD under the same set of parameters. For continuous responses, Turner et al\cite{turner_sample_2014} derived closed-form sample size formulae in traditional TSPDs, where the total sample size needed to detect a treatment effect of size $\Delta \tau$, a selection effect of size $\Delta \nu$, and a preference effect of size $\Delta \pi$, in a fixed sample TSPD are respectively given by 
$$N_{\tau}=\frac{4 \sigma_c^{2}\left(z_{1-\beta}+z_{1-{\alpha}/{2}}\right)^{2}}{(1-\theta) \Delta \tau^{2}}$$
$$N_{\nu}=\frac{\left(z_{1-\beta}+z_{1-{\alpha}/{2}}\right)^{2}}{4 \theta \phi^{2}(1-\phi)^{2} \Delta \nu^{2}}\left[\sigma_c^{2}+\phi(1-\phi)\left((2 \phi-1) \Delta \nu+\Delta \pi\right)^{2}+2\left(\frac{\theta}{1-\theta}\right)\left(\phi^{2}+(1-\phi)^{2}\right) \sigma_c^{2}\right]$$
$$N_{\pi}= \frac{\left(z_{1-\beta}+z_{1-{\alpha}/{2}}\right)^{2}}{4 \theta \phi^{2}(1-\phi)^{2} \Delta \pi^{2}}\left[\sigma_c^{2}+\phi(1-\phi)\left((2 \phi-1) \Delta \pi+\Delta \nu\right)^{2}+2\left(\frac{\theta}{1-\theta}\right)\left(\phi^{2}+(1-\phi)^{2}\right) \sigma_c^{2}\right].$$
Note that the sample size for the treatment effect is independent of the other two effects, whereas the sample sizes for selection and preference effects require assumptions about the other's effect size. We can also use the \verb|preference| package in R to calculate the required sample size for TSPDs.\cite{cameron_preference_2020} We can obtain the maximum sample size needed for the GS-TSPD by multiplying the fixed TSPD sample size by an inflation factor. Same as testing the treatment effect in group sequential RCTs, the inflation factors for Pocock and OBF alpha-spending functions can be found in published tables (such as those found in Jennison and Turnbull\cite{jennison_group_1999}) or directly from statistical software such as the \verb|gsDesign| R package, \cite{anderson_gsdesign_2022} which requires specification of the participant recruitment pattern, the maximum number of interim analyses, the alpha-spending function, the overall type I error rate, and the power.

\subsection{Extensions of group sequential two-stage preference design to accommodate binary responses} \label{subsection:binar}

The above work for the continuous outcome can be extended to a binary outcome, based on the linear probability model \eqref{eq: binomial model}. The test statistics of analysis for the $l^{th}$ interim analysis follow the same form as Equations \eqref{eq: treatment effect test stats}-\eqref{eq: preference effect test stats} with the variance expressions  calculated based on the binomial distribution. These variances were first derived by Cameron et al.,\cite{cameron_extensions_2018} and are presented in Appendix 1 for reference.

The total sample size needed to detect a treatment effect of size $\Delta \tau$, a selection effect of size $\Delta \nu$, or a preference effect of size $\Delta \pi$ for a fixed sample binary TSPD are specified as follows:\cite{cameron_extensions_2018}
$$
N_{\tau}=\frac{2\left(p_{1}\left(1-p_{1}\right)+p_{2}\left(1-p_{2}\right)\right)}{(1-\theta) \Delta \tau^{2}}\left(z_{1-{\alpha}/{2} }+z_{1-\beta}\right)^{2} 
$$
\begin{align*}
N_{\nu} = & \frac{\left(z_{1-{\alpha}/{2}}+z_{1-\beta}\right)^{2}}{4 \theta \Delta \nu^{2} \phi^{2}(1-\phi)^{2}}\left[ \phi p_{11}\left(1-p_{11}\right)+(1-\phi) p_{22}\left(1-p_{22}\right)+\frac{\left(\phi^{2} d_{1}+(1-\phi)^{2} d_{2}\right)^{2}}{\phi(1-\phi)}\right.\\
&\left.+2\left(\frac{\theta}{1-\theta}\right)\left(\phi^{2} p_{1}\left(1-p_{1}\right)+(1-\phi)^{2} p_{2}\left(1-p_{2}\right)\right)\right]
\end{align*}

\begin{align*}
N_{\pi} = &\frac{\left(z_{1-{\alpha}/{2}}+z_{1-\beta}\right)^{2}}{4 \theta \Delta \pi^{2} \phi^{2}(1-\phi)^{2}}\left[ \phi p_{11}\left(1-p_{11}\right)+(1-\phi) p_{22}\left(1-p_{22}\right)+\frac{\left(\phi^{2} d_{1}-(1-\phi)^{2} d_{2}\right)^{2}}{\phi(1-\phi)}\right.\\ &\left.+2\left(\frac{\theta}{1-\theta}\right)\left(\phi^{2} p_{1}\left(1-p_{1}\right)+(1-\phi)^{2} p_{2}\left(1-p_{2}\right)\right)\right],
\end{align*}
where $d_1=p_{11}-p_1$ and $d_2=p_{22}-p_2$; $p_i$ is the proportion of patients in the random arm responding to treatment $i$ and $p_{i i}$ is the proportion of patients in the choice arm responding to treatment $i$, analagous to the mean response parameters defined in Table \ref{tab:definitions}. As with the continuous outcome, the maximum GS-TSPD sample size is obtained by multiplying the fixed TSPD sample size by an inflation factor, which can be found in published tables\cite{jennison_group_1999} or through the \verb|gsDesign| package.\cite{anderson_gsdesign_2022} Moreover, all other aspects of GS-TSPD design and analysis are similar to those outlined in sections \ref{subsection:seq} and \ref{subsection:samplesize}. 

\section{Simulation studies} \label{section:sims}
\subsection{Simulation design}
We conducted a series of simulation studies to investigate the properties of the proposed GS-TSPD with continuous and binary responses. For all simulations, we set the nominal type I error rate to be $\alpha = 0.05$, power to be $1-\beta = 90\%$, and the maximum number of analyses to be $L = 3$. We assumed equally-spaced interim analyses and equal allocation to treatments 1 and 2 in the random arm for all scenarios considered. We conducted $10,000$ simulations and calculated the average group sequential sample size under Pocock and OBF boundaries, and the empirical power and type I error rates. All simulations were carried out using R version 3.6.2.\cite{Cite_R} 

In each scenario, we varied the preference rate for treatment $1$ ($\phi$), the first-stage randomization ratio ($\theta$), the true treatment difference ($\Delta\tau$), true selection difference ($\Delta\nu)$, and true preference difference ($\Delta\pi)$. Based on these parameters, we calculated the fixed sample TSPD and maximum GS-TSPD sample sizes according to the specific alpha-spendng function (Pocock or OBF) and the primary effect of interest (treatment, selection, or preference). The group sequential test statistics of analysis of the simulated data were calculated based on Equations \eqref{eq: treatment effect test stats}-\eqref{eq: preference effect test stats}. The test statistics of analysis were then compared to the boundary statistics given by the pre-specified alpha-spending function. The stopping sample sizes were recorded at the first interim analysis where the stopping boundaries were exceeded, and the average for the scenario was taken over all $10,000$ iterations. Simulations were conducted for both continuous and binary outcomes. R code for implementing the numerical experiments and replicating the following simulation results is available in the supplemental materials as well as at the GitHub repository: \url{https://github.com/Ruyi-Liu/GSTSPD}.

\begin{table}[htbp]
\centering
\caption{Sample size, empirical type I error and empirical power estimates for the treatment effect ($\Delta\tau$) in a group sequential two-stage preference design (GS-TSPD) for nominal type I error rate $\alpha = 0.05$, $90$\% nominal power, true selection difference $\Delta \nu = 0.2$, true preference difference $\Delta \pi = 0.3$, and  maximum number of analyses $L=3$. $\phi$ is the preference for treatment $1$ in the choice arm; and $\theta$ is the proportion randomized to the choice arm in the first stage. ``TSPD'' denotes the required sample size for a fixed sample TSPD, ``Avg GS-TSPD'' denotes the mean of the GS-TSPD stopping sample size for the Pocock and O'Brien Fleming (OBF) alpha-spending function, and ``ratio'' denotes the average GS-TSPD sample size to fixed TSPD sample size ratio. ``SD (GS-TSPD)'' represents the standard deviation of the GS-TSPD stopping sample size. Results are from $10,000$ simulations.\label{T1}}
\label{table: continuous treatment}
\begin{tabular}{rrrrrrrrrrrrrr}
    \toprule
    \multirow{2}{*}{$\phi$} & \multirow{2}{*}{$\theta$}& \multirow{2}{*}{$\Delta \tau$} & \multirow{2}{*}{\bf TSPD} &
    \multicolumn{2}{c}{\bf Avg GS-TSPD}&\multicolumn{2}{c}{\bf Ratio}&\multicolumn{2}{c}{\bf SD (GS-TSPD)}&\multicolumn{2}{c}{\bf Type I error}&\multicolumn{2}{c}{\bf Power}\\ 
    \cmidrule(lr){5-6} \cmidrule(lr){7-8} \cmidrule(lr){9-10} \cmidrule(lr){11-12} \cmidrule(lr){13-14}
     & & & & Pocock & OBF & Pocock & OBF & Pocock & OBF & Pocock & OBF & Pocock & OBF\\
    \midrule
        0.3 & 0.5 & 0.2 & 2102 & 1543 & 1687 & 0.734 & 0.803 & 651 & 419 & 0.050 & 0.050 & 0.893 & 0.896 \\ 
        ~ & ~ & 0.4 & 526 & 382 & 420 & 0.726 & 0.798 & 164 & 108 & 0.050 & 0.052 & 0.889 & 0.893 \\
        ~ & ~ & 0.6 & 234 & 171 & 187 & 0.731 & 0.799 & 73 & 48 & 0.057 & 0.053 & 0.895 & 0.895 \\\smallskip
        ~ & ~ & 0.8 & 132 & 95 & 105 & 0.720 & 0.795 & 41 & 28 & 0.059 & 0.056 & 0.896 & 0.899 \\
        0.5 & 0.5 & 0.2 & 2102 & 1547 & 1694 & 0.736 & 0.806 & 655 & 421 & 0.051 & 0.052 & 0.892 & 0.892 \\ 
        ~ & ~ & 0.4 & 526 & 384 & 422 & 0.730 & 0.802 & 162 & 107 & 0.050 & 0.051 & 0.897 & 0.893 \\
        ~ & ~ & 0.6 & 234 & 170 & 187 & 0.726 & 0.799 & 73 & 49 & 0.052 & 0.052 & 0.889 & 0.891 \\ \smallskip
        ~ & ~ & 0.8 & 132 & 96 & 105 & 0.727 & 0.795 & 41 & 28 & 0.058 & 0.054 & 0.899 & 0.898 \\
        0.7 & 0.5 & 0.2 & 2102 & 1519 & 1685 & 0.723 & 0.802 & 646 & 413 & 0.048 & 0.051 & 0.889 & 0.888 \\
        ~ & ~ & 0.4 & 526 & 385 & 422 & 0.732 & 0.802 & 163 & 105 & 0.056 & 0.053 & 0.897 & 0.892 \\
        ~ & ~ & 0.6 & 234 & 171 & 186 & 0.731 & 0.795 & 72 & 48 & 0.057 & 0.053 & 0.891 & 0.896 \\\smallskip
        ~ & ~ & 0.8 & 132 & 95 & 105 & 0.720 & 0.795 & 41 & 28 & 0.059 & 0.057 & 0.899 & 0.899 \\ 
        0.5 & 0.3 & 0.2 & 1502 & 1092 & 1208 & 0.727 & 0.804 & 463 & 299 & 0.051 & 0.051 & 0.890 & 0.892 \\ 
        ~ & ~ & 0.4 & 376 & 272 & 302 & 0.723 & 0.803 & 116 & 76 & 0.050 & 0.049 & 0.894 & 0.893 \\ 
        ~ & ~ & 0.6 & 167 & 120 & 134 & 0.719 & 0.802 & 52 & 34 & 0.056 & 0.052 & 0.897 & 0.893 \\\smallskip
        ~ & ~ & 0.8 & 94 & 68 & 75 & 0.723 & 0.798 & 29 & 20 & 0.060 & 0.058 & 0.903 & 0.901 \\ 
        0.5 & 0.7 & 0.2 & 3503 & 2546 & 2819 & 0.727 & 0.805 & 1084 & 693 & 0.048 & 0.048 & 0.893 & 0.892 \\
        ~ & ~ & 0.4 & 876 & 641 & 703 & 0.732 & 0.803 & 272 & 178 & 0.052 & 0.051 & 0.896 & 0.899 \\
        ~ & ~ & 0.6 & 390 & 281 & 312 & 0.721 & 0.800 & 121 & 80 & 0.055 & 0.055 & 0.895 & 0.891 \\
        ~ & ~ & 0.8 & 219 & 158 & 174 & 0.721 & 0.795 & 68 & 46 & 0.060 & 0.056 & 0.901 & 0.898 \\ 
    \bottomrule
\end{tabular}
\end{table}

\subsection{Continuous responses}
Considering the continuous response model \eqref{eq: normal continuous model}, we varied the true test effect size for the treatment $\Delta \tau$, selection  $\Delta \nu$, and preference $\Delta \pi$ effects from $0.2$ to $0.8$ in increments of $0.2$. 
We set the overall mean response $\mu$ to $0$, and generated the random Normal error term as $\epsilon_{ijk} \sim N(0,1)$.  
We considered scenarios where the first-stage randomization ratio for the choice arm was $\theta \in \{0.3, 0.5, 0.7\}$ and the treatment $1$ preference rate was $\phi \in \{0.3, 0.5, 0.7\}$. At each interim analysis, the number of newly-enrolled participants randomized to the choice arm choosing treatment $1$ followed a Binomial distribution with probability $\phi$, such that  $m_1^{(l)} \sim \text{Binomial}(m^{(l)},\phi)$. 

\begin{table}[htbp]
\centering
\caption{Sample size, empirical type I error and empirical power estimates for the selection effect ($\Delta\nu$) in a group sequential two-stage preference design (GS-TSPD) for nominal type I error rate $\alpha = 0.05$, $90$\% nominal power, true treatment difference $\Delta \tau = 0.4$, true preference difference $\Delta \pi = 0.3$, and maximum number of analyses $L=3$. $\phi$ is the preference for treatment $1$ in the choice arm; and $\theta$ is the proportion randomized to the choice arm in the first stage. ``TSPD'' denotes the required sample size for a fixed sample TSPD, ``Avg GS-TSPD'' denotes the mean of the GS-TSPD stopping sample size for the Pocock and O'Brien Fleming (OBF) alpha-spending function, and ``ratio'' denotes the average GS-TSPD sample size to fixed TSPD sample size ratio. ``SD (GS-TSPD)'' represents the standard deviation of the GS-TSPD stopping sample size. Results are from $10,000$ simulations.\label{T1}}
\label{table: continuous selection}
\begin{tabular}{rrrrrrrrrrrrrrr}
    \toprule
    \multirow{2}{*}{$\phi$} & \multirow{2}{*}{$\theta$}& \multirow{2}{*}{$\Delta \nu$} & \multirow{2}{*}{\bf TSPD} &
    \multicolumn{2}{c}{\bf Avg GS-TSPD}&\multicolumn{2}{c}{\bf Ratio}&\multicolumn{2}{c}{\bf SD (GS-TSPD)}&\multicolumn{2}{c}{\bf Type I error}&\multicolumn{2}{c}{\bf Power}\\ 
    \cmidrule(lr){5-6} \cmidrule(lr){7-8} \cmidrule(lr){9-10} \cmidrule(lr){11-12} \cmidrule(lr){13-14}
     & & & & Pocock & OBF & Pocock & OBF & Pocock & OBF & Pocock & OBF & Pocock & OBF\\
    \midrule
    0.3 & 0.5 & 0.2 & 6464 & 4691 & 5182 & 0.726 & 0.802 & 1997 & 1287 & 0.051 & 0.051 & 0.893 & 0.895 \\ 
        ~ & ~ & 0.4 & 1612 & 1175 & 1294 & 0.729 & 0.803 & 501 & 321 & 0.052 & 0.050 & 0.889 & 0.891 \\ 
        ~ & ~ & 0.6 & 716 & 526 & 581 & 0.735 & 0.811 & 222 & 139 & 0.051 & 0.052 & 0.882 & 0.882 \\ \smallskip
        ~ & ~ & 0.8 & 403 & 304 & 329 & 0.754 & 0.816 & 126 & 79 & 0.053 & 0.051 & 0.874 & 0.871 \\ 
        0.5 & 0.5 & 0.2 & 4251 & 3076 & 3410 & 0.724 & 0.802 & 1309 & 835 & 0.049 & 0.048 & 0.899 & 0.896 \\
        ~ & ~ & 0.4 & 1063 & 776 & 862 & 0.730 & 0.811 & 327 & 211 & 0.050 & 0.048 & 0.887 & 0.890 \\ 
        ~ & ~ & 0.6 & 473 & 351 & 385 & 0.742 & 0.814 & 146 & 92 & 0.046 & 0.048 & 0.880 & 0.884 \\ \smallskip
        ~ & ~ & 0.8 & 266 & 202 & 220 & 0.759 & 0.827 & 82 & 51 & 0.046 & 0.048 & 0.870 & 0.868 \\ 
        0.7 & 0.5 & 0.2 & 6524 & 4718 & 5237 & 0.723 & 0.803 & 2001 & 1289 & 0.051 & 0.051 & 0.896 & 0.901 \\
        ~ & ~ & 0.4 & 1642 & 1196 & 1319 & 0.728 & 0.803 & 507 & 323 & 0.046 & 0.055 & 0.896 & 0.888 \\ 
        ~ & ~ & 0.6 & 736 & 544 & 600 & 0.739 & 0.815 & 228 & 145 & 0.052 & 0.054 & 0.884 & 0.889 \\ \smallskip
        ~ & ~ & 0.8 & 418 & 317 & 342 & 0.758 & 0.818 & 130 & 81 & 0.047 & 0.051 & 0.869 & 0.871 \\ 
        0.5 & 0.3 & 0.2 & 5083 & 3689 & 4067 & 0.726 & 0.800 & 1576 & 1011 & 0.054 & 0.050 & 0.895 & 0.898 \\
        ~ & ~ & 0.4 & 1271 & 925 & 1024 & 0.728 & 0.806 & 390 & 251 & 0.050 & 0.053 & 0.889 & 0.888 \\ 
        ~ & ~ & 0.6 & 565 & 417 & 462 & 0.738 & 0.818 & 174 & 110 & 0.049 & 0.053 & 0.887 & 0.891 \\ \smallskip
        ~ & ~ & 0.8 & 318 & 241 & 261 & 0.758 & 0.821 & 99 & 61 & 0.049 & 0.050 & 0.874 & 0.879 \\ 
        0.5 & 0.7 & 0.2 & 5038 & 3663 & 4053 & 0.727 & 0.804 & 1560 & 989 & 0.047 & 0.050 & 0.890 & 0.893 \\ 
        ~ & ~ & 0.4 & 1260 & 928 & 1021 & 0.737 & 0.810 & 389 & 249 & 0.051 & 0.047 & 0.886 & 0.888 \\ 
        ~ & ~ & 0.6 & 560 & 422 & 456 & 0.754 & 0.814 & 175 & 110 & 0.050 & 0.053 & 0.873 & 0.881 \\
        ~ & ~ & 0.8 & 315 & 242 & 259 & 0.768 & 0.822 & 99 & 62 & 0.052 & 0.051 & 0.863 & 0.865 \\
    \bottomrule
\end{tabular}
\end{table}

\subsubsection{Average group sequential sample size, type I error and power}
Tables \ref{table: continuous treatment}-\ref{table: continuous preference} present the empirical type I error, empirical power, and the sample size estimate results of the GS-TSPD for detecting treatment, selection, and preference effect, respectively. Across all scenarios, the empirical power is close to the nominal value of $90$\%. The empirical type I error rates are well controlled at the nominal level in all scenarios where a nontrivial number of participants are observed in each arm. Given a large treatment effect size (e.g., $\Delta \tau = 0.8$ in Table \ref{table: continuous treatment}), the empirical type I error rates may be slightly inflated, possibly due to smaller maximum sample size and hence even smaller sample size during each interim analysis. For all three effects, the Pocock average group sequential sample size is always smaller than the average group sequential sample size under the OBF boundary. In addition, they are both smaller than the required fixed TSPD sample size under the same power, indicating that the group sequential method does, on average, save resources regardless of the type of effect being tested. Although the Pocock alpha-spending approach provides a smaller average sample size, its variation in stopping sample size appears always larger than the OBF design. In other words, using the Pocock boundary, we have a less precise estimate of what the average stopping size would be compared to the OBF boundary as it leads to a wider confidence interval for the average stopping sample size, but the Pocock boundary on average gives a smaller stopping sample size.

When testing for the selection or preference effect, our results show that, as the expected preference rate or first-stage randomization ratio increases from $0$ to $1$, the average GS-TSPD sample size first decreases and then increases. A graphical example in Appendix 3 demonstrates how the average sample size for detecting specific selection effect changes in a quadratic fashion with the increase of first-stage randomization ratio $\theta$. The change in average sample size is symmetric, with the minimum around a randomization ratio of $0.5$. This pattern holds for testing the preference effect as well. Thus, to design the most efficient GS-TSPD, the first stage randomization should be close to $\theta =0.5$; this is in accordance to previously published results for optimal allocation in fixed sample TSPDs.\cite{walter_optimal_2012}

\begin{table}[htbp]
\centering
\caption{Sample size, empirical type I error and empirical power estimates  for the  preference effect ($\Delta\pi$) in a group sequential two-stage preference design (GS-TSPD) for nominal type I error rate $\alpha = 0.05$, $90$\% nominal power, true treatment difference $\Delta \tau = 0.4$, true selection difference $\Delta \nu = 0.2$, and maximum number of analysis $L=3$. $\phi$ is the preference for treatment $1$ in the choice arm; and $\theta$ is the proportion randomized to the choice arm in the first stage. ``TSPD'' denotes the required sample size for a fixed sample TSPD, ``Avg GS-TSPD'' denotes the mean of the GS-TSPD stopping sample size for the Pocock and O'Brien Fleming (OBF) alpha-spending function, and ``ratio'' denotes the average GS-TSPD sample size to fixed TSPD sample size ratio. ``SD (GS-TSPD)'' represents the standard deviation of the GS-TSPD stopping sample size. Results are from $10,000$ simulations. 
\label{T1}}
\label{table: continuous preference}
\begin{tabular}{rrrrrrrrrrrrrr}
    \toprule
    \multirow{2}{*}{$\phi$} & \multirow{2}{*}{$\theta$}& \multirow{2}{*}{$\Delta \pi$} & \multirow{2}{*}{\bf TSPD} &
    \multicolumn{2}{c}{\bf Avg GS-TSPD}&\multicolumn{2}{c}{\bf Ratio}&\multicolumn{2}{c}{\bf SD (GS-TSPD)}&\multicolumn{2}{c}{\bf Type I error}&\multicolumn{2}{c}{\bf Power}\\ 
    \cmidrule(lr){5-6} \cmidrule(lr){7-8} \cmidrule(lr){9-10} \cmidrule(lr){11-12} \cmidrule(lr){13-14}
     & & & & Pocock & OBF & Pocock & OBF & Pocock & OBF & Pocock & OBF & Pocock & OBF\\
    \midrule
     0.3 & 0.5 & 0.2 & 6443 & 4673 & 5155 & 0.725 & 0.800 & 1993 & 1275 & 0.051 & 0.049 & 0.895 & 0.889 \\
        ~ & ~ & 0.4 & 1609 & 1178 & 1296 & 0.732 & 0.805 & 496 & 318 & 0.046 & 0.056 & 0.889 & 0.892 \\ 
        ~ & ~ & 0.6 & 715 & 527 & 578 & 0.737 & 0.808 & 222 & 140 & 0.051 & 0.049 & 0.892 & 0.889 \\ \smallskip
        ~ & ~ & 0.8 & 403 & 300 & 329 & 0.744 & 0.816 & 124 & 80 & 0.049 & 0.056 & 0.873 & 0.874 \\ 
        0.5 & 0.5 & 0.2 & 4224 & 3065 & 3384 & 0.726 & 0.801 & 1299 & 843 & 0.047 & 0.051 & 0.893 & 0.898 \\ 
        ~ & ~ & 0.4 & 1056 & 775 & 855 & 0.734 & 0.810 & 327 & 207 & 0.047 & 0.048 & 0.893 & 0.891 \\ 
        ~ & ~ & 0.6 & 470 & 350 & 383 & 0.745 & 0.815 & 145 & 92 & 0.053 & 0.051 & 0.883 & 0.882 \\ \smallskip
        ~ & ~ & 0.8 & 264 & 199 & 219 & 0.754 & 0.830 & 82 & 51 & 0.051 & 0.049 & 0.877 & 0.872 \\ 
        0.7 & 0.5 & 0.2 & 6483 & 4684 & 5191 & 0.723 & 0.801 & 1998 & 1288 & 0.051 & 0.051 & 0.897 & 0.897 \\ 
        ~ & ~ & 0.4 & 1629 & 1191 & 1309 & 0.731 & 0.804 & 503 & 322 & 0.048 & 0.048 & 0.892 & 0.893 \\ 
        ~ & ~ & 0.6 & 729 & 540 & 592 & 0.741 & 0.812 & 225 & 143 & 0.049 & 0.052 & 0.884 & 0.884 \\ \smallskip
        ~ & ~ & 0.8 & 413 & 311 & 337 & 0.753 & 0.816 & 129 & 81 & 0.053 & 0.054 & 0.873 & 0.875 \\ 
        0.5 & 0.3 & 0.2 & 5039 & 3650 & 4043 & 0.724 & 0.802 & 1551 & 980 & 0.052 & 0.048 & 0.899 & 0.897 \\ 
        ~ & ~ & 0.4 & 1260 & 921 & 1015 & 0.731 & 0.806 & 387 & 247 & 0.052 & 0.048 & 0.895 & 0.894 \\ 
        ~ & ~ & 0.6 & 560 & 416 & 455 & 0.743 & 0.813 & 174 & 108 & 0.050 & 0.050 & 0.886 & 0.890 \\ \smallskip
        ~ & ~ & 0.8 & 315 & 239 & 259 & 0.759 & 0.822 & 98 & 60 & 0.051 & 0.047 & 0.877 & 0.876 \\
        0.5 & 0.7 & 0.2 & 5019 & 3646 & 4025 & 0.726 & 0.802 & 1558 & 1000 & 0.053 & 0.053 & 0.894 & 0.895 \\ 
        ~ & ~ & 0.4 & 1255 & 921 & 1008 & 0.734 & 0.803 & 388 & 249 & 0.051 & 0.052 & 0.882 & 0.886 \\ 
        ~ & ~ & 0.6 & 558 & 412 & 454 & 0.738 & 0.814 & 174 & 111 & 0.051 & 0.054 & 0.878 & 0.877 \\ 
        ~ & ~ & 0.8 & 314 & 240 & 258 & 0.764 & 0.822 & 98 & 62 & 0.052 & 0.052 & 0.860 & 0.866 \\ 
    \bottomrule
\end{tabular}
\end{table}

\begin{figure}[htbp]
    \centering
    \subfigure{\includegraphics[width=0.7\textwidth]{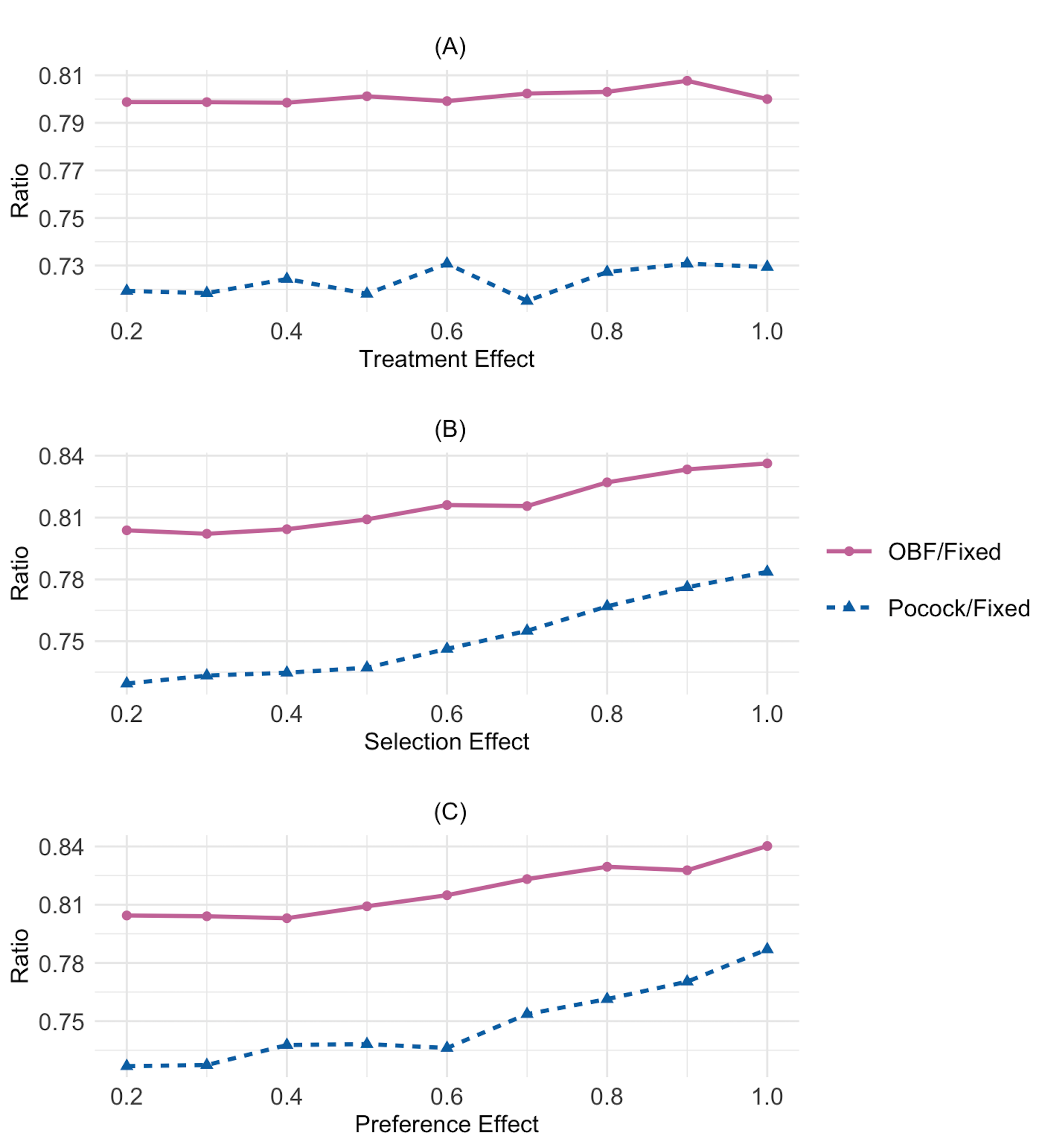}} 
    \caption{Change in average sample size ratio by test effect size. The average group sequential two-stage preference design (GS-TSPD) sample size to fixed TSPD sample size ratios are calculated from $10,000$ simulations under the following conditions: proportion of individuals in the choice arm choosing treatment $1$ $\phi = 0.5$, first stage allocation ratio $\theta = 0.5$, Type I error rate $\alpha = 0.05$, $90$\% power, and maximum number of analyses $L=3$. The treatment effect is varied in panel (A) with fixed selection ($\Delta \nu = 0.2$) and preference ($\Delta \pi = 0.3$) effects. The selection effect is varied in panel (B) with fixed treatment ($\Delta \tau = 0.4$) and preference ($\Delta \pi = 0.3$) effects. The preference effect is varied in panel (C) with fixed treatment ($\Delta \tau = 0.4$) and selection ($\Delta \nu = 0.2$) effects.}
    \label{fig: sample size ratio}
\end{figure}

\subsubsection{Sample size ratio}
Tables \ref{table: continuous treatment}-\ref{table: continuous preference} also present the ratio between the average GS-TSPD sample size and fixed TSPD sample size to determine the proportion of expected sample size savings using the group sequential approach. We found that the ratio is generally invariant to the expected preference rate $\phi$ and the first-stage randomization ratio $\theta$. However, as can be seen in Figure \ref{fig: sample size ratio} when we varied the test effect size (holding all else constant), we found that the sample size ratio for the fixed sample TSPD and GS-TSPD still remains invariant if we test for treatment effect. However, if we test for the selection or preference effect, the ratio increases with the test effect size. For example, the benefit of GS-TSPD over the fixed sample TSPD is on average a $20$\% savings when using the OBF boundary and approximately a $28$\% savings when using the Pocock boundary if the treatment effect is of primary interest regardless of the hypothesized treatment effect size. But there appear to be less savings for the selection and preference effects when using the GS-TSPD for detecting larger effect sizes. To understand the reason behind this trend, we calculated the empirical probability of stopping at each analysis, which we present in Appendix 4. Since the selection and preference effects have symmetric test statistics, the empirical findings on the probability of stopping are comparable and we only present the selection effect results. We see that, as the alternative increases, the probability of stopping at earlier analyses decreases for both the Pocock and OBF boundaries; that is, the probability of stopping at later analyses or reaching the final analysis increases. This may be partially due to the fact that the variance of the preference and selection effects are a function of the preference and selection effect sizes, such that increases in effect size necessitate larger variances; this is unique from the treatment effect, whose estimate is independent from its asymptotic variance and shifts in magnitude of the treatment effect can be made while variance remains constant.

\subsection{Binary responses}
Considering the binary response model \eqref{eq: binomial model}, we varied the treatment, selection, and preference differences from $0.1$ to $0.3$ to ensure that the proportion of participants responding to treatment $i$ in the random arm, $p_i$, and the proportion of participants responding to treatment $i$ in the choice arm, $p_{ii}$, are valid, thus smaller than $1$ and larger than $0$. We also set the overall response proportion to be $0.5$. The scenario design should ensure that the number of participants assigned to each arm at each analysis is not excessively small.\cite{cameron_extensions_2018}

\begin{table}[!htbp]
\centering
\caption{Sample size, empirical type I error and empirical power estimates for the treatment ($\Delta\tau$), selection ($\Delta\nu$) and preference effects ($\Delta\pi$) for a group sequential two-stage preference design (GS-TSPD) with a binary response for nominal type I error rate $\alpha = 0.05$, $90$\% nominal power, proportion preferring treatment $1$ in the choice arm $\phi = 0.5$, proportion allocated to the choice arm $\theta = 0.5$, and maximum number of analyses $L=3$. When testing $\Delta\tau$, $\Delta \nu =\Delta \pi = 0.1$; when testing  $\Delta\nu$, $\Delta \tau = \Delta \pi = 0.1$; and when testing $\Delta\pi$, $\Delta \tau =\Delta \nu = 0.1$. ``TSPD'' denotes the required sample size for a fixed sample TSPD, ``Avg GS-TSPD'' denotes the mean of the GS-TSPD stopping sample size for the Pocock and O'Brien Fleming (OBF) alpha-spending functions, and ``ratio'' denotes the average GS-TSPD sample size to fixed TSPD sample size ratio. ``SD (GS-TSPD)'' represents the standard deviation of the GS-TSPD stopping sample size. Results are from $10,000$ simulations.  
 \label{T1}}
\label{table: binary outcomes}
\begin{tabular}{rrrrrrrrrrrrrr}
    \toprule
    \multirow{2}{*}{\bf Test}& \multirow{2}{*}{\bf Size} & \multirow{2}{*}{\bf TSPD} &
    \multicolumn{2}{c}{\bf Avg GS-TSPD}&\multicolumn{2}{c}{\bf Ratio}&\multicolumn{2}{c}{\bf SD (GS-TSPD)}&\multicolumn{2}{c}{\bf Type I error}&\multicolumn{2}{c}{\bf Power}\\ 
    \cmidrule(lr){4-5} \cmidrule(lr){6-7} \cmidrule(lr){8-9} \cmidrule(lr){10-11} \cmidrule(lr){12-13}
    & & & Pocock & OBF & Pocock & OBF & Pocock & OBF & Pocock & OBF & Pocock & OBF\\
    \midrule

    $\Delta \tau$ & 0.10 & 2082 & 1485 & 1661 & 0.713 & 0.798 & 646 & 410 & 0.052 & 0.051 & 0.899 & 0.897 \\
    ~ & 0.15 & 914  & 650  & 727  & 0.711 & 0.795 & 285 & 185 & 0.054 & 0.051 & 0.896 & 0.898 \\
    ~ & 0.20 & 506  & 368  & 399  & 0.727 & 0.789 & 157 & 103 & 0.052 & 0.057 & 0.899 & 0.905 \\ 
    ~ & 0.25 & 316  & 230  & 250  & 0.728 & 0.791 & 98  & 67  & 0.054 & 0.055 & 0.901 & 0.902 \\
    ~ & 0.30 & 214  & 152  & 167  & 0.710  & 0.780  & 66  & 49  & 0.060  & 0.058 & 0.895 & 0.911 \\ 

    \midrule
    $\Delta \nu$ & 0.10 & 4098 & 2959 & 3257 & 0.722 & 0.795 & 1262 & 820 & 0.049 & 0.048 & 0.898 & 0.900 \\
    ~ & 0.15 & 1801 & 1297 & 1436 & 0.720 & 0.797 & 552 & 361 & 0.051 & 0.054 & 0.899 & 0.901 \\ 
    ~ & 0.20 & 999 & 716 & 802 & 0.717 & 0.803 & 306 & 198 & 0.050 & 0.047 & 0.900 & 0.898 \\ 
    ~ & 0.25 & 628 & 453 & 503 & 0.721 & 0.801 & 193 & 126 & 0.052 & 0.052 & 0.899 & 0.898 \\ 
    ~ & 0.30 & 428 & 309 & 343 & 0.722 & 0.801 & 131 & 85 & 0.047 & 0.055 & 0.891 & 0.900 \\ 
    
    \midrule
    $\Delta \pi$ & 0.10 & 4098 & 2935 & 3278 & 0.716 & 0.800 & 1256 & 821 & 0.051 & 0.048 & 0.901 & 0.903 \\ 
        ~ & 0.15 & 1810 & 1306 & 1452 & 0.722 & 0.802 & 559 & 357 & 0.053 & 0.048 & 0.899 & 0.900 \\ 
        ~ & 0.20 & 1009 & 730 & 806 & 0.723 & 0.799 & 311 & 200 & 0.049 & 0.046 & 0.890 & 0.897 \\ 
        ~ & 0.25 & 639 & 462 & 510 & 0.723 & 0.798 & 197 & 127 & 0.051 & 0.051 & 0.893 & 0.896 \\ 
        ~ & 0.30 & 437 & 316 & 348 & 0.723 & 0.796 & 135 & 88 & 0.051 & 0.054 & 0.899 & 0.898 \\

    \bottomrule
\end{tabular}
\end{table}

Table \ref{table: binary outcomes} summarizes the sample size properties of our proposed GS-TSPD for binary outcomes, as well as the empirical power and type I error rates. Similar to the continuous results, the average GS-TSPD sample size to fixed TSPD sample size ratio does not appear to interact with the expected preference rate and the first-stage randomization ratio, so we only present results where $\theta = 0.5$ and $\phi = 0.5$. The type I error rates for testing treatment, selection, or preference effects are well controlled in the designed scenarios with sufficient sample size in each arm, which suggests that the Normal approximation holds well for all three effects when the sample size in each arm and the response rate are moderate. The empirical power is also well preserved in each design scenario. Also, similar to the continuous outcome results, the Pocock average sample size is always smaller than the OBF average sample size, and they are both smaller than the required fixed TSPD sample size under the same power. Concordant with the simulations with a continuous response, the Pocock sample size usually has a larger standard deviation and appears more variable from simulation iteration to iteration.

\section{Illustrative Data Examples} 
\label{section:example}
We further illustrate our proposed GS-TSPD framework in the context of a comparative care delivery trial for HCV. The primary aim of the study is to determine differences in patients receiving HCV care through two different delivery methods (speciality clinic or mobile medical clinic (MMC)). The hypotheses are two-fold: first, we hypothesize that those who receive care at MMCs will have better outcomes (a larger increase in HRQoL as measured by the mental component summary (MCS) and better HCV cure rates); second, we hypothesize that those receiving their preferred mode of delivery may have better health outcomes.\cite{ware_12-item_1996}

\subsection{Health-related quality of life}
We first demonstrate the applicability of the GS-TSPD for continuous outcomes using the MCS of the HRQoL. We assume the MMC preference rate is $0.56$, the true treatment, selection, and preference effects are $\Delta {\tau} = -2$, $\Delta {\nu} = -2.1$, and $\Delta {\pi} = 3.9$, respectively, and the variance of the MCS is $\sigma_c^2=16$. These parameters are obtained from Cameron and Esserman\cite{cameron_sample_2018}, except that we consider smaller effect sizes for the selection and preference effects for better illustration of the potential sample size reduction using GS-TSPD. In addition, although we have multiple hypotheses corresponding to the different effects, in this data example we first illustrate how one might plan a study where one of the treatment, selection, or preference effects is alone the primary objective of interest. Assuming a type I error rate of $0.05$ and a power of $90$\%, the sample sizes for a GS-TSPD with $L=3$ analyses and a fixed sample TSPD required to detect a treatment effect of size $-2$, a selection effect of size $-2.1$, and a preference effect of size $3.9$ are presented in the first three rows of Table \ref{table: motivating example}. Note that these calculations are carried for each effect separately.

\begin{table}[htbp]
\centering
\caption{Required sample sizes for detecting treatment ($N_{\tau}$), selection ($N_{\nu}$), and preference effects ($N_{\pi}$) for health related quality of life (HRQoL) and Hepatitis C virus (HCV) cure using fixed sample two-stage preference design (TSPD) and group sequential (GS)-TSPD for HCV patients receiving care at either a specialty clinic or mobile medical clinic. HRQoL and ultimate cure rate are different endpoints under which the required sample size calculations are subject to different design parameters, including true effect size. For HRQoL, we assume the following: treatment effect $\Delta {\tau} = -2$, selection effect $\Delta {\nu} = -2.1$, and preference effect $\Delta {\pi} = 3.9$.  For HCV cure rate, we assume: $\Delta {\tau} = 0.15$, $\Delta {\nu} = -0.185$, and $\Delta {\pi} = 0.297$. There are at most $L=3$ analyses. We set the type I error rate to $0.05$ and the power to $90$\%. \label{T1}}
\label{table: motivating example}
\begin{tabular}{rrrrrrr}
    \toprule
    \multirow{2}{*}{\bf Outcome }& \multirow{2}{*}{\bf Effect} & \multirow{2}{*}{\bf Fixed Design} &\multicolumn{2}{c}{\bf Average GS-TSPD}&\multicolumn{2}{c}{\bf Maximum GS-TSPD}\\ 
    \cmidrule(lr){4-5} \cmidrule(lr){6-7} 
     & & & Pocock & OBF & Pocock & OBF\\
    \midrule

    \multirow{3}{*}{HRQoL}& Treatment ($N_{\tau}$) & 337& 268 & 286 & 388 & 343\\
   
    & Selection ($N_{\nu}$) &  697 & 531 & 577 & 802 & 708\\
    
    & Preference ($N_{\pi}$) & 188 & 146 & 157& 216& 191\\

    \midrule
    \multirow{3}{*}{Cure Rate}& Treatment ($N_{\tau}$) & 764& 546 & 610 & 880 & 777\\
   
    & Selection ($N_{\nu}$) & 863 & 628 & 688 & 993& 877\\
    
    & Preference ($N_{\pi}$) & 322 & 234 & 260 & 371& 328\\
    
    \bottomrule
\end{tabular}
\end{table} 

To detect the treatment effect of size $-2$, the fixed sample TSPD would need $337$ participants, whereas the GS-TSPD with Pocock and OBF boundaries would require an average of $268$ and $286$ participants and a maximum of $388$ and $343$ participants, respectively. Compared to the fixed sample TSPD, our proposed GS-TSPD results in a $15$\% reduction in sample size on average. Similar results are seen for the selection and preference effects. To detect the selection effect of size $-2.1$, the fixed sample TSPD would need $697$ participants, whereas the GS-TSPD with Pocock and OBF boundaries would require an average of $531$ and $577$ participants and a maximum of $802$ and $708$ participants, respectively. To detect the preference effect of size $3.9$, the fixed sample TSPD would need $188$ participants, whereas the GS-TSPD with Pocock and OBF boundaries would require an average of $146$ and $157$ participants and a maximum of $216$ and $191$ participants, respectively. Although the maximum sample sizes under the GS-TSPDs are larger than the comparable fixed sample TSPD, we see that the GS-TSPDs may result in a smaller average stopping sample size. In addition, particularly under the OBF design, the maximum sample size required under the GS-TSPD may only be trivially larger than the required sample size under the fixed design. In this example, the design using an OBF boundary only requires a $2$\% increase of the maximum sample size over the comparable fixed sample TSPD ($343$ total participants versus $337$ needed for the fixed sample design).

If we were to power the overall study based only on the preference effect hypothesis that those receiving their preferred mode of delivery will have a larger increase in HRQoL, the GS-TSPD with Pocock and OBF boundaries requires $146$ and $157$ participants on average. However, in this case, the other effects of secondary interest (treatment and selection) are severely underpowered and would limit the study's ability to draw inference on these points. Due to a substantial discrepancy between the sample size requirements, choosing the largest sample size among the three effects (selection effect) would overpower the study for the preference effect -- making it possible for us to have larger than required power to detect a clinically meaningful preference effect size -- but enable us to also detect the effects of secondary interest with sufficient power. To mitigate this concern, one may consider the framework of hierarchical testing, where methods for multiplicity adjustment with respect to power and strong control of the familywise Type I error rate for both primary and secondary hypotheses have been previously investigated.\cite{glimm_hierarchical_2010} We do not pursue this framework here but refer to Section \ref{section:discussion} for a discussion. 

\subsection{HCV cure rate}
As an additional illustration, we estimate the sample sizes needed for testing the impact of care delivery on HCV cure for the treatment, selection, and preference effects. We assume that the preference rate for MMC is $0.56$, and the true treatment, selection, and preference effects are respectively $\Delta {\tau} = 0.15$, $\Delta {\nu} = -0.185$, and $\Delta {\pi} = 0.297$. These design parameters are obtained from Cameron et al. in the context of TSPD with a binary outcome.\cite{cameron_extensions_2018} Then, we calculate the necessary sample size for the corresponding fixed sample TSPD and $L=3$ GS-TSPD to respectively detect the treatment effect, selection effect, and preference effect sizes with $90$\% power. The results are presented in the bottom three rows of Table \ref{table: motivating example}. From Table \ref{table: motivating example}, to detect the treatment effect of $0.15$, a fixed sample TSPD requires $764$ patients, whereas the GS-TSPD with Pocock and OBF boundaries requires an average of $546$ and $610$ patients and a maximum of $880$ and $777$ patients, respectively. Thus, group sequential monitoring reduces the sample size by $20$\% on average. Similarly, to detect the selection effect of $-0.185$, a fixed sample TSPD requires $863$ patients, whereas the GS-TSPD with Pocock and OBF boundaries requires an average of $628$ and $688$ patients and a maximum of $993$ and $877$ patients, respectively; to detect the preference effect of $0.297$, a fixed sample TSPD requires $322$ patients, whereas the GS-TSPD with Pocock and OBF boundaries requires an average of $234$ and $260$ patients and a maximum of $371$ and $328$ patients, respectively. This again indicates that applying the group sequential method to TSPD can be more cost efficient. 

\section{Discussion} \label{section:discussion}
As a pragmatic trial design, a TSPD is attractive because it not only examines the efficacy of a treatment in the study population but also accounts for estimation of potential selection and preference effects. To potentially save design resources without compromising the study power, this work considers applying a group sequential monitoring framework to TSPD, which gradually adds participants to the study and decides to terminate or continue the trial based on repeated significance tests on accumulating data.\cite{armitage_repeated_1969} From this standpoint, this work presents a possible solution to one of the potential limitations for adopting a traditional TSPD, namely the relatively large sample size required to detect the selection and preference effects. We presented the group sequential test statistics for treatment, selection, and preference effect. Binary responses were modeled using a linear probability model, and a Normal approximation was applied for resulting test statistics in GS-TSPD. We considered a linear probability model for binary outcomes rather than a logistic regression as we are interested in the effects on the risk difference scale rather than the multiplicative scale; the linear probability model performs well considering our target estimands and their direct interpretability.\cite{gomila_logistic_2021} To ensure that there is no deterministic preference for treatment and that we can obtain a sufficiently large sample from the population preferring each of the two treatments, preliminary work or a pilot study should be pursued prior to a full TSPD or GS-TSPD to investigate the potential preference rate for each treatment; this may prevent trials where preference for a particular treatment is rare. Furthermore, we discussed how to approximate the sequential sampling density and obtain stopping boundaries given by Pocock and OBF alpha-spending functions by verifying the independent increments assumption necessary to apply an Armitage approximation to the sequential densities. We also discussed how to compute the maximum GS-TSPD sample size under fixed statistical power, as well as the average GS-TSPD sample size required to detect each effect at the design stage. In simulations, we assumed uniform accrual, where patients' entries at each interim analysis are equal-sized, and focused on short-term endpoints, meaning that the outcomes for participants are available by the time of each interim analysis; thus, we do not need to consider proxy endpoints in this simple setting. If the accrual is not uniform (i.e., unequal numbers of patients are accrued in equal amounts of calendar time), the calendar-timing of interim analyses should be adjusted to accommodate this.\cite{emerson_frequentist_2007} If the outcome in a pragmatic trial is measured over a relatively long period of time compared to the calendar timing of the interim analyses, we may use the short-term proxy outcome or partial information in the analyses.\cite{niewczas2019interim}

Across all three effect quantities, our results showed that the average GS-TSPD sample size can be potentially smaller than the fixed TSPD sample size while controlling the type I error rate and preserving statistical power. Therefore, our proposed GS-TSPD framework represents a practical strategy to address one possible barrier for testing selection and preference effects when the study resources are constrained. We saw that the Pocock boundary resulted in a smaller expected sample size, though with less precision, than the OBF boundary. On the contrary, we observed that the Pocock boundary resulted in a larger maximum sample size than the OBF boundary due to the less precise stopping sample size. For continuous outcomes, when testing for selection or preference effect, the proportion of samples saved by GS-TSPD on average decreased with an increasing standardized effect size. Interestingly, we saw that a GS-TSPD is more advantageous when the investigator is interested in a small selection or preference effect, for which it would normally be difficult or infeasible to recruit sufficient participants in a fixed sample TSPD. The same analysis could not be performed in the binary outcome scenarios due to the difficulty with the explicit mean-variance relationship posed to standardizing the effect size. We also saw that the Normal approximation of selection and preference effect estimators for binary responses maintained desired operating characteristics when sample sizes and overall response proportion are moderate.

In this article we only considered studies designed using Pocock and OBF alpha-spending functions, though many others exist with varying design characteristics. In the course of choosing a particular alpha-spending function to use to design a study, it is helpful to consider not only the average sample size but also the empirical probability of stopping at each analysis. Appendix 4 presents the stopping probabilities at each of three interim analyses for testing a selection effect under Pocock and OBF boundaries. By comparing these two sets of boundaries, we can assess whether the probability of requiring more participants than the fixed sample TSPD, i.e., the probability of proceeding to the final analysis and enrolling the maximum sample size, is more acceptable in one case compared to the other. In these examples, the probability of reaching the maximum sample size is around $30$\% for the Pocock boundaries while it is around $50$\% for the OBF boundaries. If the investigator still considers these probabilities too large, they may explore other boundary shapes, number of interim analyses, and analysis timing that may lower this probability to a level they consider more reasonable, though these modifications may also alter other design aspects like maximum sample size. Note that, in this paper, we focused on efficacy stopping rules, but the same principle could be applied to develop a stopping procedure for futility by using type II error spending functions; under futility stopping, the trial will terminate early and no additional data will be collected if the treatment signal is so small such that one will be unlikely to reject the null hypothesis.\cite{pampallona1995spending,pampallona_group_1994}

We intend to address several limitations of our work in future investigations. To begin with, we assumed a constant preference rate across the entire study population. However, in reality, the proportion of patients preferring a given treatment could vary between subgroups or across time, and additional research is needed to understand how these non-constant preference rates may affect GS-TSPD operating characteristics. In addition, throughout this work we assumed there were no undecided patients. Further work is needed to extend our GS-TSPD design to these more complicated scenarios, where the undecided patients go through the second-stage randomization.\cite{walter_optimising_2019}

As we have seen, GS-TSPD can be an additional tool to be considered for investigators who are interested in testing selection or preference effects but are discouraged by the potentially large sample sizes required by a fixed sample TSPD. Many trials are sized based on a minimum meaningful effect size, even if a larger, more optimistic effect size is anticipated. Thus, it is imperative that the planned total sample size should be sufficiently large to power in the case of minimum meaningful effect, but the interim monitoring component of GS-TSPD can plan for the possibility of early termination if the observed effect size is more optimistic. However, it is not intended to be applicable to every pragmatic study. A GS-TSPD design may not be sufficiently beneficial if investigators expect data to be accrued very quickly, meaning that interim analyses would need to take place very close together in calendar time, creating logistical issues. To apply GS-TSPD in a pragmatic trial, the investigators should take into account the effect size, estimated accrual speed, endpoint type, type of alpha-spending functions, and maximum number of interim analyses to ensure that investigators are able and willing to recruit to the corresponding maximum GS-TSPD sample size when necessary. By approximating the average sample size and empirical probability of stopping at each analysis, the investigators may determine whether utilizing GS-TSPD may bring sufficient benefits compared to a fixed sample TSPD. With these factors being considered, the potential reduction in sample size and the increased speed to conclude the study can serve as motivators for conducting a GS-TSPD.

We also note that our proposed GS-TSPD assumes testing is being conducted on only a single primary endpoint. Powering a study based only on the primary effect of interest may underpower or overpower the other effects of secondary interest if there is a large discrepancy in estimated sample sizes. The potential reconcilability of this discrepancy in sample size should be comprehensively examined by the investigative team. For instance, in cases where sample sizes exhibit minimal disparities, researchers may opt to slightly overpower some hypotheses while slightly underpowering other hypotheses. It is imperative to emphasize that the viability of such reconciliation varies depending on the research team and the specific research objectives, necessitating a comprehensive examination within the context of each individual study and context. A study underpowered for secondary effects runs the risk of investigators being unable to draw helpful inference about secondary hypotheses, while a study overpowered for secondary effects make it possible to conclude statistical significance for effect sizes that are too small to be clinically meaningful. It is also possible that an investigator may be equally interested in two or more endpoints\cite{tang_design_1989} as co-primary outcomes.\cite{asakura_interim_2017} An intersection-union testing framework within the sequential monitoring procedure could accommodate this, but warrants a separate development as the multiple endpoints are correlated, meaning that the joint density must be re-derived, and the independent increments assumption cannot be directly utilized. Luo and Quan\cite{luo2023some} have recently investigated multiplicity adjustment procedures with multiple endpoints and multiple interim analyses, which may be adapted to the GS-TSPD. Furthermore, even with a single primary outcome, it may be relevant to address the challenge in simultaneously testing for multiple hypotheses (such as whether any of treatment, selection, or preference effects is equal to $0$), which may require modifications to the stopping boundaries and evaluation of the familywise error rate and power properties after controlling for multiplicity.\cite{tamhane_group_2021,rosenblum_multiple_2016}

\ack
\section*{Acknowledgements}
Research in this article was funded by CTSA Grant Number UL1 TR001863 from the National Center for Advancing Translational Science (NCATS), a component of the National Institutes of Health (NIH). The statements presented in this article are solely the responsibility of the authors and do not necessarily represent the views of the National Institutes of Health.

\newpage
\section*{Appendix 1. Calculation of estimated variance in test statistics}
The test statistics for selection and preference effects include estimated variances in the denominators, i.e. $\widehat{\operatorname{Var}}\left(z_{1}^{(l*)}-z_{2}^{(l*)}\right)$ and $\widehat{\operatorname{Var}}\left(z_{1}^{(l*)}+z_{2}^{(l*)}\right)$, where $z_1^{(l*)} = \left(\sum_{h=1}^{m_1^{(l*)}}x_{1h}^{(l*)}\right) - m_1^{(l*)}\overline{y_1^{(l*)}}$ and $z_2^{(l*)} = \left(\sum_{h=1}^{m_2^{(l*)}}x_{2h}^{(l*)}\right) - m_2^{(l*)}\overline{y_2^{(l*)}}$. In order to obtain the explicit formulas for the test statistics, we can break the estimated variances into three parts:
$$\widehat{\operatorname{Var}}\left(z_{1}^{(l*)} \pm z_{2}^{(l*)}\right)=\widehat{\operatorname{Var}}\left(z_{1}^{(l*)}\right)+\widehat{\operatorname{Var}}\left(z_{2}^{(l*)}\right) \pm 2 \widehat{\operatorname{Cov}}\left(z_{1}^{(l*)}, z_{2}^{(l*)}\right) $$
The estimators for each of the variances can be expressed as:
$$
\begin{aligned}
\widehat{\operatorname{Var}}\left(z_{i}^{(l*)}\right)&=m_{i}^{(l*)} \widehat{\operatorname{Var}}\left(x_{i}^{(l*)}\right)+\left(1+\frac{m^{(l*)}-1}{m^{(l*)}} m_{i}^{(l*)}\right) m_{i}^{(l*)} \frac{\widehat{\operatorname{Var}}\left(y_{i}^{(l*)}\right)}{n_{i}^{(l*)}}\\
&+\frac{m_{1}^{(l*)} m_{2}^{(l*)}}{m^{(l*)}}\left(\overline{x_{i}^{(l*)}}-\overline{y_{i}^{(l*)}}\right)^{2}, \text{ for } i=1,2 
\end{aligned}$$
$$\widehat{\operatorname{Cov}}\left(z_{1}^{(l*)}, z_{2}^{(l*)}\right)=-\frac{m_{1}^{(l*)} m_{2}^{(l*)}}{m^{(l*)}}\left(\overline{x_{1}^{(l*)}}-\overline{y_{1}^{(l*)}}\right)\left(\overline{x_{2}^{(l*)}}-\overline{y_{2}^{(l*)}}\right)$$
For continuous responses, $\widehat{\operatorname{Var}}\left(x_{i}^{(l*)}\right)$ and $\widehat{\operatorname{Var}}\left(y_{i}^{(l*)}\right)$ are estimated by the empirical variance in each cell. For binary responses, the variances follow the standard form of a Bernoulli random variable:
$$\widehat{\operatorname{Var}}\left(x_{i}^{(l*)}\right) = \overline{x_{i}^{(l*)}}\left(1-\overline{x_{i}^{(l*)}}\right)$$
$$\widehat{\operatorname{Var}}\left(y_{i}^{(l*)}\right) = \overline{y_{i}^{(l*)}}\left(1-\overline{y_{i}^{(l*)}}\right)$$

\section*{Appendix 2. Proof for independent increments}
First recall that the cumulative test statistic for the selection difference is:
$$\quad S_{l, \nu} = \sum_{r=1}^{l}{\mathcal{T}}_{\nu}(r) = \sum_{r=1}^{l}\frac{z_{1}^{(r)}-z_{2}^{(r)}}{\sqrt{\widehat{\operatorname{Var}}\left(z_{1}^{(r)}-z_{2}^{(r)}\right)}}. $$

To show that the cumulative test statistics for selection difference has independent increments, we want to show  
$$\operatorname{Cov}\left(S_{l-1,\nu},S_{l,\nu} -S_{l-1,\nu} \right) = \operatorname{Cov}\left(\sum_{r=1}^{l-1} {\mathcal{T}}_{\nu}(r),  {\mathcal{T}}_{\nu}(l)\right) = \operatorname{Cov}\left( {\mathcal{T}}_{\nu}(1),  {\mathcal{T}}_{\nu}(l)\right) + \cdots + \operatorname{Cov}\left( {\mathcal{T}}_{\nu}(l-1),  {\mathcal{T}}_{\nu}(l)\right)=0$$ 
Thus, it is sufficient to verify independent increments assumption if we prove  $\operatorname{Cov}\left( {\mathcal{T}}_{\nu}(s),  {\mathcal{T}}_{\nu}(t)\right) = 0$
for any $s, t \in \{1,..., L\}, t \neq s$. Considering the selection effect test statistics, it is equivalent to show  $\operatorname{Cov}(z_{1}^{(t)}-z_{2}^{(t)},z_{1}^{(s)}-z_{2}^{(s)}) = 0$ for any $s, t \in \{1,..., L\}, t \neq s$.

We have $m_1^{(t)}$ and $m_1^{(s)}$ are independent, where $m_1^{(t)}\sim Binomial(m^{(t)},\phi)$ and $m_1^{(s)}\sim Binomial(m^{(s)},\phi)$. It then follows that $\mathbb{E}(m_1^{(t)}) =m^{(t)}\phi$, $Var(m_1^{(t)})= m^{(t)}\phi(1-\phi)$ for all $t \in \{1,...,L\}$. Also let $x_{1}^{(t)}$ be independent of $x_{1}^{(s)}$ and $x_{2}^{(s)}$ for any $s \neq t$, $y_1^{(t)}$ be independent of $y_1^{(s)}$ and $y_2^{(s)}$ for any $s \neq t$, and $y$ be independent of $m_1$ and $x$ within and between each interim analysis. Based on these assumptions, we can express $\operatorname{Cov}(z_{1}^{(t)}-z_{2}^{(t)},z_{1}^{(s)}-z_{2}^{(s)})$ as:
$$
\begin{aligned}
        \operatorname{Cov}(z_{1}^{(t)}-z_{2}^{(t)},z_{1}^{(s)}-z_{2}^{(s)}) &= \operatorname{Cov} \left(\left(\sum_{r=1}^{m_{1}^{(t)}} x_{1 r}^{(t)}\right)-m_{1}^{(t)} \overline{{y}_{1}^{(t)}} -\left(\sum_{r=1}^{m_{2}^{(t)}} x_{2 r}^{(t)}\right)+m_{2}^{(t)} \overline{y_{2}^{(t)}} \right.,\\
        & \left.\left(\sum_{r=1}^{m_{1}^{(s)}} x_{1 r}^{(s)}\right)-m_{1}^{(s)} \overline{y_{1}^{(s)}} -\left(\sum_{r=1}^{m_{2}^{(s)}} x_{2 r}^{(s)}\right)+m_{2}^{(s)} \overline{y_{2}^{(s)}} \right)\\
\end{aligned}
$$
(1) To find this covariance, we must compute the pairwise covariances between each of the two sets of four components, making 16 covariances in total. First, we compute the covariances between $\left(\sum_{r=1}^{m_{1}^{(t)}} x_{1 r}^{(t)}\right)$ in the first set and the four components in the second set:
$$
\begin{aligned}
&\operatorname{Cov} \left(\sum_{r=1}^{m_{1}^{(t)}} x_{1 r}^{(t)},\sum_{r=1}^{m_{1}^{(s)}} x_{1 r}^{(s)}\right)-\operatorname{Cov}\left(\sum_{r=1}^{m_{1}^{(t)}} x_{1 r}^{(t)}, m_{1}^{(s)} \overline{{y}_{1}^{(s)}}\right)-\operatorname{Cov}\left(\sum_{r=1}^{m_1^{(t)}} x_{1 r}^{(t)},\sum_{r=1}^{m_{2}^{(s)}} x_{2 r}^{(s)}\right)\\
& +\operatorname{Cov}\left(\sum_{r=1}^{m_{1}^{(t)}} x_{1 r}^{(t)}, m_{2}^{(s)} \overline{y_{2}^{(s)}}\right)
\end{aligned}
$$

(a)
$$
\begin{aligned}
\operatorname{Cov}\left(\sum_{r=1}^{m_{1}^{(t)}} x_{1 r}^{(t)}, \sum_{r=1}^{m_{1}^{(s)}} x_{1 r}^{(s)} \right) &= \mathbb{E}\left(\left(\sum_{r=1}^{m_{1}^{(t)}} x_{1 r}^{(t)}\right) \left(\sum_{r=1}^{m_{1}^{(s)}} x_{1 r}^{(s)}\right)\right)\\
&-\mathbb{E}\left(\sum_{r=1}^{m_{1}^{(t)}} x_{1 r}^{(t)}\right)\mathbb{E}\left(\sum_{r=1}^{m_{1}^{(s)}} x_{1 r}^{(s)}\right)\\
\left(\text{by law of total expectation}\right) &= \mathbb{E}\left[\mathbb{E}\left(\left(\sum_{r=1}^{m_{1}^{(t)}} x_{1 r}^{(t)}\right) \left(\sum_{r=1}^{m_{1}^{(s)}} x_{1 r}^{(s)}\right) \middle\vert\ m_1^{(t)},m_1^{(s)}\right)\right]\\
& - \mathbb{E}(m_1^{(t)})\mathbb{E}(x_1^{(t)})\mathbb{E}(m_1^{(s)})\mathbb{E}(x_1^{(s)})\\
& =\mathbb{E}\left[\mathbb{E}\left(\sum_{r=1}^{m_{1}^{(t)}} x_{1 r}^{(t)} \middle\vert\ m_1^{(t)}\right) \mathbb{E}\left(\sum_{r=1}^{m_{1}^{(s)}} x_{1 r}^{(s)} \middle\vert\ m_1^{(s)}\right)\right]\\
& -m^{(t)} \phi \cdot \mu_{11} \cdot m^{(s)} \phi \cdot \mu_{11}\\
& = \mathbb{E}\left[\left(\sum_{r=1}^{m_{1}^{(t)}}\mathbb{E}\left( x_{1 r}^{(t)} \middle\vert\ m_1^{(t)}\right)\right) \cdot \left(\sum_{r=1}^{m_{1}^{(s)}}\mathbb{E}\left( x_{1 r}^{(s)} \middle\vert\ m_1^{(s)}\right)\right)\right]\\
& -m^{(t)} \phi \cdot \mu_{11} \cdot m^{(s)} \phi \cdot \mu_{11} \\
\left(\text{response variables }x_{11}^{(t)},\cdots , x_{1m_{1}^{(t)}}^{(t)}\text{ are i.i.d} \right)& =\mathbb{E}\left(m_1^{(t)} \cdot \mu_{11} \cdot m_1^{(s)} \mu_{11} \right) -m^{(t)} \phi \cdot \mu_{11} \cdot m^{(s)} \phi \cdot \mu_{11}\\
& = \mu_{11}^2 \mathbb{E}\left(m_1^{(t)}\right) \cdot \mathbb{E}\left(m_1^{(s)}\right) -m^{(t)} \phi \cdot \mu_{11} \cdot m^{(s)} \phi \cdot \mu_{11}\\
& = \mu_{11}^2 m^{(t)}\phi \cdot m^{(s)}\phi -m^{(t)} \phi \cdot \mu_{11} \cdot m^{(s)} \phi \cdot \mu_{11}\\
& = 0
\end{aligned}
$$

(b)   
$$
\begin{aligned}
\operatorname{Cov}\left(\sum_{r=1}^{m_{1}^{(t)}} x_{1r}^{(t)}, \sum_{r=1}^{m_{2}^{(s)}} x_{2 r}^{(s)}\right) &=\mathbb{E}\left(\sum_{r=1}^{m_{1}^{(t)}} x_{1r}^{(t)} \cdot \sum_{r=1}^{{m_{2}^{(s)}}} x_{2 r}^{(s)}\right)-\mathbb{E}\left(\sum_{r=1}^{m_{1}^{(t)}} x_{1r}^{(t)}\right) \mathbb{E}\left(\sum_{r=1}^{m_{2}^{(s)}} x_{2 r}^{(s)}\right) \\
&=\mu_{11} \mu_{22} \cdot \mathbb{E}\left(m_{1}^{(t)}\right) \mathbb{E}\left(m^{(s)}-m_{1}^{(s)}\right)-m^{(t)} \phi(m^{(s)}-m^{(s)} \phi) \cdot \mu_{11} \mu_{22} \\
&=\mu_{11} \mu_{22} m^{(t)} \phi(m^{(s)}-m^{(s)} \phi)-m^{(t)} \phi(m^{(s)}-m^{(s)} \phi) \mu_{11} \mu_{22}\\
&=0
\end{aligned}
$$

(c)   
$$
\begin{aligned}
\operatorname{Cov}\left(\sum_{r=1}^{m_{1}^{(t)}} x_{1 r}^{(t)}, m_{1}^{(s)} \overline{y_{1}^{(s)}}\right) & = \operatorname{Cov}\left(\sum_{r=1}^{m_{1}^{(t)}} x_{1 r}^{(t)}, \frac{m_{1}^{(s)}}{n_{1}^{(s)}} \sum_{h=1}^{n_{1}^{(s)}} y_{1 h}^{(s)}\right) \\
= & \frac{1}{n_{1}^{(s)}}\left(\mathbb{E}\left(\sum_{r=1}^{m_{1}^{(t)}} x_{1 r}^{(t)} \cdot m_{1}^{(s)} \sum_{h=1}^{n_{1}^{(s)}} y_{1h}^{(s)}\right)-\mathbb{E}\left(\sum_{r=1}^{m_{1}^{(t)}} x_{1 r}^{(t)}\right) \mathbb{E}\left(m_{1}^{(s)} \sum_{h=1}^{n_{1}^{(s)}} y_{1h}^{(s)}\right)\right)\\
\left(\text{Note that } y \independent m_1, y \independent x \right) =&\frac{1}{n_{1}^{(s)}}\mathbb{E}\left(\sum_{r=1}^{m_{1}^{(t)}} x_{1 r}^{(t)} \cdot m_{1}^{(s)}\right) \mathbb{E}\left(\sum_{h=1}^{n_{1}^{(s)}} y_{1h}^{(s)}\right)-\mathbb{E}\left(m_{1}^{(t)}\right) \mathbb{E}\left(x_{1}^{(t)}\right)  \mathbb{E}\left(m_{1}^{(s)}\right)  \cdot \mu_{1}\\
=& \mathbb{E}\left(\sum_{r=1}^{m_{1}^{(t)}} x_{1 r}^{(t)} \cdot m_{1}^{(s)}\right) \cdot \mu_{1}-m^{(t)}m^{(s)} \phi^{2} \cdot \mu_{11} \mu_{1}\\
=& \sum_{b=1}^{m^{(s)}}b \cdot\mathbb{E}\left(\sum_{r=1}^{m_1^{(t)}}x_{1r}^{(t)}\right) \mathbb{P}\left(m_1^{(s)} = b\right) \cdot \mu_{1}-m^{(t)}m^{(s)} \phi^{2} \cdot \mu_{11} \mu_{1} \\
\left(m_1^{(t)},m_1^{(s)} \text{ i.i.d}\right)=& \mathbb{E}\left(m_{1}^{(t)}\right) \mathbb{E}\left(m_{1}^{(s)}\right) \cdot \mu_{11} \mu_{1}-m^{(t)}m^{(s)} \phi^{2} \cdot \mu_{11} \mu_{1} \\
=& m^{(t)}m^{(s)} \phi^{2} \cdot \mu_{11} \mu_{1}-m^{(t)}m^{(s)} \phi^{2} \cdot \mu_{11} \mu_{1} \\
=& 0
\end{aligned}
$$

(d)  
$$
\begin{aligned}
\operatorname{Cov}\left(\sum_{r=1}^{m_{1}^{(t)}} x_{1 r}^{(t)}, m_{2}^{(s)} \overline{{y}_{2}^{(s)}}\right)&=\mathbb{E}\left(\sum_{r=1}^{m_{1}^{(t)}} x_{1 r}^{(t)} \cdot \frac{m_{2}^{(s)}}{n_{2}^{(s)}} \sum_{h=1}^{n_{2}^{(s)}} y_{2h}^{(s)}\right)-\mathbb{E}\left(\sum_{r=1}^{m_{1}^{(t)}} x_{1 r}^{(t)}\right) \cdot \frac{1}{n_{2}^{(s)}} \cdot \mathbb{E}\left(m_2^{(s)} \sum_{h=1}^{n_{2}^{(s)}} y_{2h}^{(s)}\right) \\
&= \frac{1}{n_{2}^{(s)}} \mathbb{E}\left(m_{2}^{(s)} \sum_{r=1}^{m_{1}^{(t)}} x_{1 r}^{(t)}\right) \mathbb{E}\left(y_{2}^{(s)}\right) \cdot n_{2}^{(s)} -\mathbb{E}\left(m_{1}^{(t)}\right) \mathbb{E}\left(x_{1}^{(t)}\right) \mathbb{E}\left(m_{2}^{(s)}\right) \mathbb{E}\left(y_{2}^{(s)}\right) \\
&= \mu_{2} \cdot \mu_{11} \cdot \mathbb{E}\left(m_{1}^{(t)}\right) \mathbb{E}\left(m^{(s)}-m_{1}^{(s)}\right)-\mathbb{E}\left(m_{1}^{(t)}\right) \cdot \mu_{11} \cdot \mu_{2}\cdot \mathbb{E}\left(m^{(s)}-m_{1}^{(s)}\right) \\
&= 0
\end{aligned}
$$
(2)  
Next, we must compute similar covariances between the second element of the first set, $m_{1}^{(t)} \overline{{y}_{1}^{(t)}}$, and the four components in the second set:
$$
\begin{aligned}
& -\operatorname{Cov}\left(m_{1}^{(t)} \overline{{y}_{1}^{(t)}}, \sum_{r=1}^{m_1^{(s)}} x_{1 r}^{(s)}\right)+\operatorname{Cov}\left(m_{1}^{(t)} \overline{{y}_{1}^{(t)}}, m_{1}^{(s)} \overline{{y}_{1}^{(s)}}\right)+\operatorname{Cov}\left(m_{1}^{(t)} \overline{{y}_{1}^{(t)}}, \sum_{r=1}^{m_{2}^{(s)}} x_{2 r}^{(s)}\right)\\
& -\operatorname{Cov}\left(m_{1}^{(t)} \overline{{y}_{1}^{(t)}}, m_{2}^{(s)} \overline{{y}_{2}^{(s)}}\right)
\end{aligned}
$$  
(a)
$$
\begin{aligned}
\operatorname{Cov}\left(m_{1}^{(t)} \overline{{y}_{1}^{(t)}}, \sum_{r=1}^{m_1^{(s)}} x_{1 r}^{(s)}\right)&=\mathbb{E}\left(\frac{m_{1}^{(t)}}{n_{1}^{(t)}} \sum_{h=1}^{n_{1}^{(t)}} y_{1 h}^{(t)} \cdot \sum_{r=1}^{m_{1}^{(s)}} x_{1 r}^{(s)}\right)-\frac{1}{n_{1}^{(t)}} \mathbb{E}\left(m_{1}^{(t)} \sum_{h=1}^{n_{1}^{(t)}} y_{1h}^{(t)}\right) \cdot \mathbb{E}\left(\sum_{r=1}^{m_{1}^{(s)}} x_{1r}^{(s)}\right)\\
&=\mathbb{E}\left(y_{1}^{(t)}\right) \cdot \mathbb{E}\left(m_{1}^{(t)} \sum_{r=1}^{m_{1}^{(s)}} x_{1 r}^{(s)}\right)-\mathbb{E}\left(m_{1}^{(t)}\right) \cdot \mathbb{E}\left(y_{1}^{(t)}\right) \cdot \mathbb{E}\left(m_{1}^{(s)}\right) \mathbb{E}\left(x_{1}^{(s)}\right)\\
&=\mu_{1} \cdot\left(\sum_{a=1}^{m^{(t)}} \mathbb{P}\left(m_{1}^{(t)}=a\right) a \cdot \sum_{b=1}^{m^{(s)}} \mathbb{P}\left(m_{1}^{(s)}=b\right) \cdot b \cdot \mathbb{E}\left(x_{1}^{(s)}\right)\right)\\
& -m^{(t)}\cdot m^{(s)} \phi^{2} \mu_{1} \mu_{11}\\
&=\mu_{1} \mathbb{E}\left(m_{1}^{(t)}\right) \mathbb{E}\left(m_{1}^{(s)}\right) \cdot \mathbb{E}\left(x_{1}^{(s)}\right)-m^{(t)}\cdot m^{(s)} \phi^{2} \mu_{1} \mu_{11}\\
&= \mu_{1} m^{(t)}\cdot m^{(s)} \phi^{2}  \mu_{11}-m^{(t)}\cdot m^{(s)} \phi^{2} \mu_{1} \mu_{11}\\
&=0
\end{aligned}
$$
(b)  
$$
\begin{aligned}
\operatorname{Cov}\left(m_{1}^{(t)} \overline{y_{1}^{(t)}}, m_{1}^{(s)} \overline{y_{1}^{(s)}}\right) &=\mathbb{E}\left(m_{1}^{(t)} \overline{y_{1}^{(t)}} m_{1}^{(s)} \overline{y_{1}^{(s)}}\right)-\mathbb{E}\left(m_{1}^{(t)} \overline{{y}_{1}^{(t)}}\right) \mathbb{E}\left(m_{1}^{(s)} \overline{y_{1}^{(s)}}\right) \\
{\left( \text{Note that } y_{1}^{(t)} \independent y_{1}^{(s)} \text{, and } m_{1}^{(t)} \independent m_{1}^{(s)}\right) } &=\mathbb{E}\left(m_{1}^{(t)}\right) \mathbb{E}\left(y_{1}^{(t)}\right) \mathbb{E}\left(m_{1}^{(s)}\right) \mathbb{E}\left(y_{1}^{(s)}\right)\\
& -\mathbb{E}\left(m_{1}^{(t)}\right) \mathbb{E}\left(y_{1}^{(t)}\right) \mathbb{E}\left(m_{1}^{(s)}\right) \mathbb{E}\left(y_{1}^{(s)}\right)\\
&=0
\end{aligned}
$$
(c)  
$$
\begin{aligned}
\operatorname{Cov}\left(m_{1}^{(t)} \overline{y_{1}^{(t)}}, \sum_{r=1}^{m_{2}^{(s)}} x_{2 r}^{(s)}\right) &=\mathbb{E}\left(m_{1}^{(t)} \overline{y_{1}^{(t)}} \sum_{r=1}^{m_{2}^{(s)}} x_{2 r}^{(s)}\right)-\mathbb{E}\left(m_{1}^{(t)} \overline{y_{1}^{(t)}}\right) \mathbb{E}\left(\sum_{r=1}^{m_{2}^{(s)}} x_{2 r}^{(s)}\right)\\
&=\mathbb{E}\left(y_{1}^{(t)}\right) \mathbb{E}\left(m_{1}^{(t)} \sum_{r=1}^{m_{2}^{(s)}} x_{2 r}^{(s)}\right)-\mathbb{E}\left(m_{1}^{(t)}\right) \mathbb{E}\left(y_{1}^{(t)}\right) \mathbb{E}\left(m_{2}^{(s)}\right) \mathbb{E}\left(x_{2}^{(s)}\right)\\
&=\mathbb{E}\left(y_{1}^{(t)}\right) \mathbb{E}\left(m_{1}^{(t)}\right) \mathbb{E}\left(m_{2}^{(s)}\right) \cdot \mathbb{E}\left(x_{2}^{(s)}\right)-\mathbb{E}\left(y_{1}^{(t)}\right) \mathbb{E}\left(m_{1}^{(t)}\right) \mathbb{E}\left(m_{2}^{(s)}\right) \mathbb{E}\left(x_{2}^{(s)}\right)\\
&=0
\end{aligned}
$$
(d)    
$$
\begin{aligned}
\operatorname{Cov}\left(m_{1}^{(t)} \overline{{y}_{1}^{(t)}}, m_{2}^{(s)} \overline{{y}_{2}^{(s)}}\right) &=\mathbb{E}\left(m_{1}^{(t)} \overline{y_{1}^{(t)}} m_{2}^{(s)} \overline{y_{2}^{(s)}}\right)-\mathbb{E}\left(m_{1}^{(t)}\overline{{y}_{1}^{(t)}}\right) \mathbb{E}\left(m_{2}^{(s)} \overline{{y}_{2}^{(s)}}\right) \\
\left(\text{Note that } y_{1}^{(t)} \independent y_{2}^{(s)} \right) &=\mathbb{E}\left(m_{1}^{(t)}\right) \mathbb{E}\left(y_{1}^{(t)}\right) \mathbb{E}\left(m_{2}^{(s)}\right) \mathbb{E}\left(y_{2}^{(s)}\right)-\mathbb{E}\left(m_{1}^{(t)}\right) \mathbb{E}\left(y_{1}^{(t)}\right) \mathbb{E}\left(m_{2}^{(s)}\right) \mathbb{E}\left(y_{2}^{(s)}\right) \\
&=0
\end{aligned}
$$
By (a), (b), (c), (d) from part (1) and part (2), we conclude that the first eight terms of the covariance are all zero. For similar reasons, the other eight terms regarding $\left(\sum_{r=1}^{m_{2}^{(t)}} x_{2 r}^{(t)}\right)$ and $m_{2}^{(t)} \overline{y_{2}^{(t)}}$ would also equal to zero. Thus, $\operatorname{Cov}(z_{1}^{(t)}-z_{2}^{(t)},z_{1}^{(s)}-z_{2}^{(s)}) = 0$, and we may conclude that the independent increments assumption holds for selection difference. As $\operatorname{Cov}(z_{1}^{(t)}+z_{2}^{(t)},z_{1}^{(s)}+z_{2}^{(s)})$ contains the same terms as $\operatorname{Cov}(z_{1}^{(t)}-z_{2}^{(t)},z_{1}^{(s)}-z_{2}^{(s)})$ only with different signs, and since all sixteen covariance terms equal to zero, we also have $\operatorname{Cov}(z_{1}^{(t)}+z_{2}^{(t)},z_{1}^{(s)}+z_{2}^{(s)}) = 0$. 

Recall that for preference difference 
$$S_{l, \pi} = \sum_{r=1}^{l}  {\mathcal{T}}_{\pi}(r) = \sum_{r=1}^{l}\frac{z_{1}^{(r)}+z_{2}^{(r)}}{\sqrt{\widehat{\operatorname{Var}}\left(z_{1}^{(r)}+z_{2}^{(r)}\right)}}.$$
To show that the cumulative test statistics for preference difference has independent increments, we want to show  
$$\operatorname{Cov}\left(S_{l-1,\pi},S_{l,\pi} -S_{l-1,\pi} \right) = \operatorname{Cov}\left(\sum_{r=1}^{l-1} {\mathcal{T}}_{\pi}(r),  {\mathcal{T}}_{\pi}(l)\right) = \operatorname{Cov}\left( {\mathcal{T}}_{\pi}(1),  {\mathcal{T}}_{\pi}(l)\right) + \cdots + \operatorname{Cov}\left( {\mathcal{T}}_{\pi}(l-1),  {\mathcal{T}}_{\pi}(l)\right)=0.$$ 
Similar to the selection difference, it is sufficient to verify independent increments assumption for preference difference if $\operatorname{Cov}\left( {\mathcal{T}}_{\pi}(s),  {\mathcal{T}}_{\pi}(t)\right) = 0$ for any $s, t \in \{1,..., L\}, s \neq t$. Since $\operatorname{Cov}(z_{1}^{(t)}+z_{2}^{(t)},z_{1}^{(s)}+z_{2}^{(s)}) = 0$ for any $s, t \in \{1,..., L\}, s \neq t$, we obtain $\operatorname{Cov}\left( {\mathcal{T}}_{\pi}(s),  {\mathcal{T}}_{\pi}(t)\right) = 0$. Thus, the independent increments assumption also holds for preference difference.

\section*{Appendix 3. Change of average GS-TSPD sample size by first-stage randomization ratio}
\setcounter{figure}{0} 
\begin{figure}[htbp]
    \centering
    \subfigure{\includegraphics[width=0.8\textwidth]{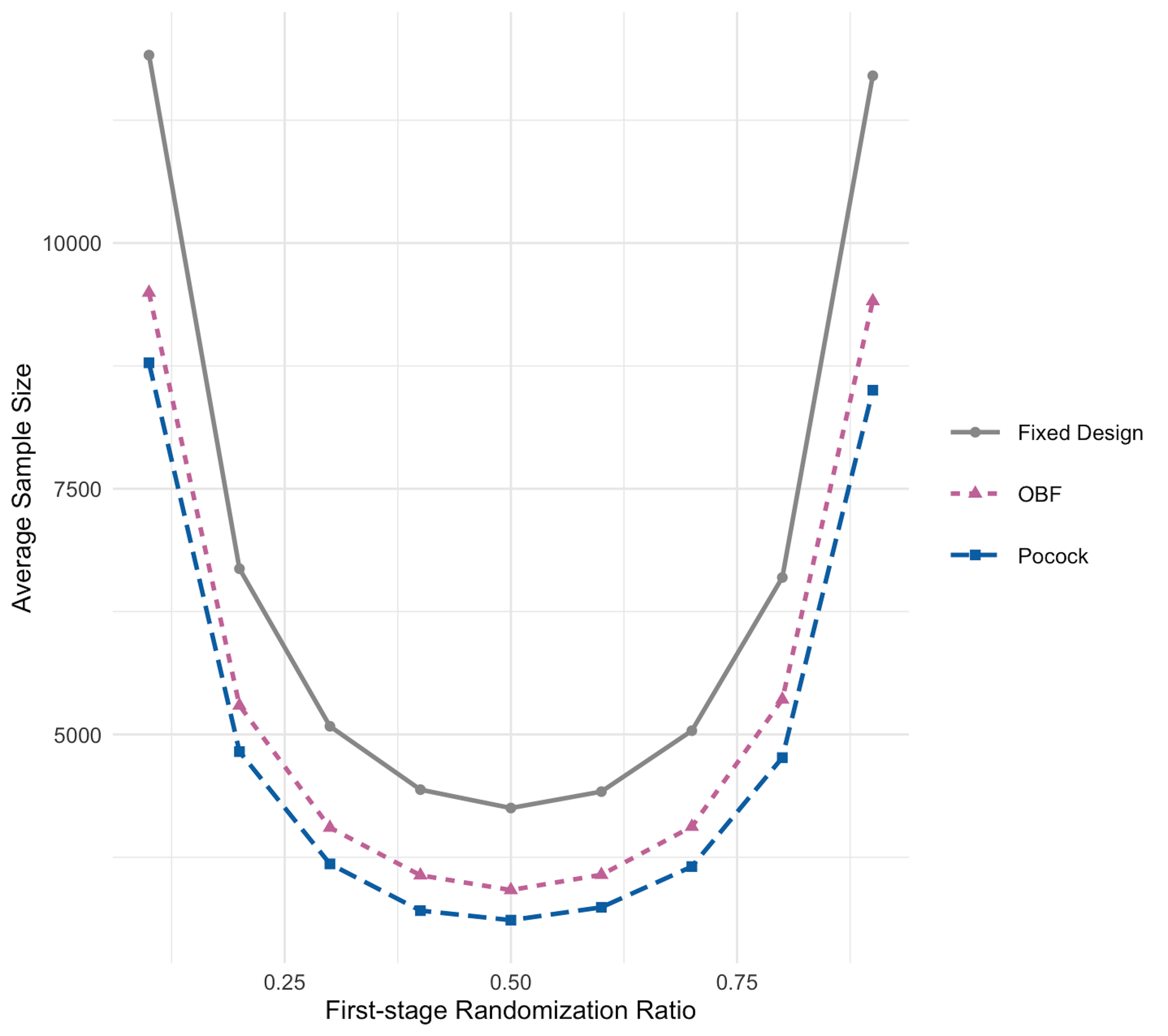}} 
    \caption{Change of average group sequential two-stage preference design (GS-TSPD) sample size by first-stage randomization ratio when detecting a selection effect of size 0.2 with a preference rate of $\phi=0.5$, assuming a nominal type I error rate $\alpha = 0.05$ and  power $1-\beta = 0.9$. We assume a true treatment difference of $\Delta \tau = 0.4$ and a true preference difference of $\Delta \pi = 0.1$, with a maximum of $L=3$ analyses.}
    \label{}
\end{figure}

\newpage
\section*{Appendix 4. Stopping probability at each analysis for detecting selection effect of different size}
\setcounter{table}{0} 
\begin{table}[!htbp]
\centering
\caption{Stopping probability at each analysis for detecting the selection effect calculated from 50000 simulations with $\phi = 0.5$, $\theta = 0.5$, $\alpha = 0.05$, $1-\beta = 0.9$, $\Delta \tau = 0.4$, $\Delta \pi = 0.3$, $L=3$.\label{T1}}
\label{table: stopping probability}
\begin{tabular}{rrrrrrrrr}
    \toprule
    \multirow{2}{*}{$\Delta \nu$} &\multicolumn{3}{c}{\bf Pocock Stopping Probability at Each Analysis}&\multicolumn{3}{c}{\bf OBF Stopping Probability at Each Analysis}\\ 
    \cmidrule(lr){2-4} \cmidrule(lr){5-7} 
     & First Analysis & Second Analysis & Third Analysis & First Analysis & Second Analysis & Third Analysis\\
    \midrule
    0.2 & 0.384 & 0.342 & 0.275 & 0.054 & 0.526 & 0.420 \\
0.3 & 0.380 & 0.343 & 0.277 & 0.054 & 0.517 & 0.429 \\
0.4 & 0.376 & 0.345 & 0.278 & 0.052 & 0.514 & 0.434 \\
0.5 & 0.362 & 0.345 & 0.293 & 0.050 & 0.509 & 0.441 \\
0.6 & 0.358 & 0.344 & 0.299 & 0.045 & 0.505 & 0.449 \\
0.7 & 0.347 & 0.341 & 0.312 & 0.042 & 0.502 & 0.456 \\
0.8 & 0.339 & 0.340 & 0.321 & 0.039 & 0.488 & 0.474 \\
0.9 & 0.328 & 0.343 & 0.329 & 0.037 & 0.482 & 0.481 \\
1.0 & 0.317 & 0.337 & 0.346 & 0.033 & 0.465 & 0.501\\
    \bottomrule
\end{tabular}
\end{table}

\renewcommand*{\thetable}{\arabic{table} }

\end{document}